\documentclass[12pt,titlepage]{article}
\usepackage{bm, amssymb,pifont,cancel, amsmath,comment,color}
\usepackage{here}
\usepackage{cite}
\usepackage{graphicx}
\usepackage{subfigure}

\makeatletter

\@addtoreset{equation}{section}
\makeatother
 
\begin{document}

\begin{titlepage}
\null
\begin{flushright}
WU-HEP-19-02\\
KEK-TH-2100
\end{flushright}

\vskip 1cm
\begin{center}
\baselineskip 0.8cm
{\LARGE \bf
Interpolation of partial and full supersymmetry breakings in $\mathcal{N}=2$ supergravity}

\lineskip .75em
\vskip 1cm

\normalsize

{\large } {\large Hiroyuki Abe} $^{1}${\def\thefootnote{\fnsymbol{footnote}}\footnote[1]{E-mail address: abe@waseda.jp}},
{\large Shuntaro Aoki} $^{1}${\def\thefootnote{\fnsymbol{footnote}}\footnote[2]{E-mail address: shun-soccer@akane.waseda.jp}}, 
{\large } {\large Sosuke Imai} $^{1}${\def\thefootnote{\fnsymbol{footnote}}\footnote[3]{E-mail address: s.i.sosuke@akane.waseda.jp}},
{\large } {\large Yutaka Sakamura} $^{2,3}${\def\thefootnote{\fnsymbol{footnote}}\footnote[4]{E-mail address: sakamura@post.kek.jp}}

\vskip 1.0em

$^1${\small\it Department of Physics, Waseda University, \\ 
Tokyo 169-8555, Japan}
\vskip 1.0em

$^2${\small{\it KEK Theory Center, Institute of Particle and Nuclear Studies, KEK,\\
1-1 Oho, Tsukuba, Ibaraki 305-0801, Japan}}
\vskip 1.0em

$^3${\small{\it Department of Particles and Nuclear Physics,\\
SOKENDAI (The Graduate University for Advanced Studies),\\
1-1 Oho, Tsukuba, Ibaraki 305-0801, Japan}}
\vskip 1.0em

\vspace{12mm}

{\bf Abstract}\\[5mm]
{\parbox{13cm}{\hspace{5mm} \small
We discuss an $\mathcal{N}=2$ supergravity model that interpolates the full and the partial supersymmetry breakings. 
In particular, we find the conditions for an $\mathcal{N}=0$ Minkowski vacuum, 
which is continuously connected to the partial-breaking ($\mathcal{N}=1$ preserving) one. 
The model contains multiple (Abelian) vector multiplets and a single hypermultiplet, 
and is constructed by employing the embedding tensor technique. 
We compute the mass spectrum on the Minkowski vacuum, and find some non-trivial mass relations among the massive fields. 
Our model allows us to choose the two supersymmetry-breaking scales independently, 
and to discuss the cascade supersymmetry breaking for the applications to particle phenomenology and cosmology. 
}}

\end{center}

\end{titlepage}

\tableofcontents
\vspace{35pt}
\hrule

\section{Introduction}\label{intro}
Extended ($\mathcal{N}\geq 2$) supergravity naturally appears from higher dimensional supergravity and string compactifications 
(see~\cite{Grana:2005jc,Blumenhagen:2006ci} for review). Its interactions are more restricted (predictive) than $\mathcal{N}=1$ supergravity, which has been intensively investigated from the viewpoint of particle phenomenology and cosmology. 
Due to the restrictions, it is a non-trivial task to obtain a phenomenologically favorable supersymmetry-breaking vacuum. 
This fact is related to the no-go theorem for the partial breaking of extended supergravity~\cite{Cecotti:1984rk, Cecotti:1984wn}. 
For example, in $\mathcal{N}=2$ supergravity, the naive gauging (electric gauging in the frame where the prepotential exists) leads to the breaking of the whole $\mathcal{N}=2$ supersymmetries,\footnote{The full supersymmetry breakings in $\mathcal{N}=2$ supergravity are also discussed based on the constrained $\mathcal{N}=2$ superfields, e.g., in Refs.~\cite{Dudas:2017sbi,Kuzenko:2017gsc}.} and some fields acquire masses by (super) higgs mechanism. 
However, it generally leads to a degenerate mass spectrum due to a single supersymmetry-breaking scale. Therefore, it is impossible to realize the above mentioned $\mathcal{N}=1$ models within this framework. In order to incorporate such phenomenological models, the $\mathcal{N}=2$ supersymmetries need to be broken by two different breaking scales. Then, an approximate $\mathcal{N}=1$ supersymmetry appears between these scales.  
Furthermore, when they are hierarchical, the situation approaches to the partial breaking of the supersymmetries.

The possibility of the partial breaking in $\mathcal{N}=2$ supergravity was found in Refs.~\cite{Ferrara:1995gu,Ferrara:1995xi,Fre:1996js}, and there, it was shown that the partial breaking occurs by gauging a matter (hyper) sector in a specific frame where the prepotential does not exist. The systematic analysis for the partial breaking conditions based on the so-called embedding tensor~\cite{deWit:2002vt,deWit:2005ub} can be found in Ref.~\cite{Louis:2009xd} (see also~\cite{Louis:2010ui,Hansen:2013dda,Antoniadis:2018blk}). These models are studied in connection with the partial breaking in the global $\mathcal{N}=2$ models~\cite{Antoniadis:1995vb} by taking the rigid supersymmetry limit~\cite{Ferrara:1995xi,David:2003dh, Andrianopoli:2015rpa,Laamara:2017hdl}. Also, there are various discussions related to the D-brane effective actions, e.g.,~\cite{Bagger:1996wp,Rocek:1997hi,GonzalezRey:1998kh,Tseytlin:1999dj,Burgess:2003hx,Antoniadis:2008uk,Kuzenko:2009ym,Ambrosetti:2009za,Kuzenko:2011ya,Ferrara:2014oka,Kuzenko:2015rfx,Ferrara:2016crd,Kuzenko:2017zla,Antoniadis:2017jsk,Farakos:2018aml,Cribiori:2018jjh}. The partial supersymmetry breaking in non-Abelian gauge theories are discussed in both of the global~\cite{Fujiwara:2005hj,Fujiwara:2005kf,Fujiwara:2006qh} and the local supersymmetry cases~\cite{Itoyama:2006ef} (see also~\cite{Maruyoshi:2006te} for review).

In this paper, we interpolate the full and the partial supersymmetry breaking to obtain the approximate partial breaking of $\mathcal{N}=2$ supersymmetries.\footnote{A similar approach in the global $\mathcal{N}=2$ supersymmetry can be found in Refs.~\cite{Antoniadis:2012cg,Laamara:2015rbq}} Then, the spectrum should be characterized by two independent breaking scales of the $\mathcal{N}=2$ supersymmetries. In Ref.~\cite{Ferrara:1995gu}, the model realizes the full or the partial supersymmetry breaking, and the parameter spaces for those vacua are continuously connected. For phenomenological applications, we need matter multiplets, in addition to the goldstino multiplets. Therefore, we generalize the model in such a way that it has multiple vector multiplets, more general prepotential, and larger class of the gauging of the isometries, including the model  of Ref.~\cite{Ferrara:1995gu} as a special case. Then, we investigate the conditions for the supersymmetry breakings at two different scales, and compute the corresponding mass spectrum. As we will see later, there are some non-trivial relations in the spectrum in this class of models.

This paper is organized as follows. In Sec.~\ref{setup}, we briefly review $\mathcal{N}=2$ supergravity and specify our model. In Sec.~\ref{Spectrum}, we derive the stationary conditions of the potential, and discuss the supersymmetry breaking by analyzing the supersymmetry transformations of the fermions. Then, we compute the boson and the fermion masses and find some relations among them. Section.~\ref{con} is devoted to the summary. In Appendix~\ref{notation}, we collect the notations. The technical details of the embedding tensor formalism are summarized in Appendix~\ref{2-form}. In Appendix~\ref{GKF}, we list the vacuum expectation values of the gauge kinetic functions. In Appendix~\ref{int}, we show the interaction terms including fermion bilinears on the vacuum, which are necessary to discuss the decay modes in Sec.~\ref{Spectrum}.

\section{Set up}\label{setup}
In this section, we specify the model which is a generalization of Ref.~\cite{Ferrara:1995gu}. Here we follow the convention of \cite{Andrianopoli:1996cm} and use the unit $M_{P}= 1$, where $M_{P}= 2.4\times 10^{18}$ GeV is the reduced Planck mass.

\subsection{Vector and hyper sectors}
We consider $\mathcal{N}=2$ gauged supergravity in four dimensions which contains $n_v$ Abelian vector multiplets, one hypermultiplet, in addition to the gravitational multiplet:  
\begin{align}
&{\rm{Vector\ multiplets:}}\ \ \{z^i, \lambda^{iA}, A^i_{\mu}\},\ \ (i=1,\cdots ,n_v)\\
&{\rm{Hypermultiplet:}}\ \ \{b^u,\zeta_{\alpha}\},\\
&{\rm{Gravitational\ multiplet:}}\ \ \{g_{\mu \nu}, \psi^A_{\mu}, A^0_{\mu}\}.
\end{align}
A vector multiplet contains a complex scalar $z^i$, two gauginos $\lambda^{iA}\ (A=1,2)$ and a vector $A^i_{\mu}$. A hypermultiplet contains four real scalars $b^u\ (u=0,\cdots,3)$\footnote{The hyperscalars $b^u\ (u=0,\cdots,3)$ have the quaternionic structure.} and two hyperinos $\zeta_{\alpha}\ (\alpha=1,2)$.\footnote{Note that $\alpha$ is not a spinor index. The spinor indices are suppressed throughout this paper.} The gravitational multiplet contains the spacetime metric $g_{\mu \nu}\ (\mu,\nu=0,\cdots,3)$, two gravitinos $\psi^A_{\mu}\ (A=1,2)$ and the graviphoton $A_{\mu}^0$. 
Note that there are $N+1$ vector fields in the system and they are labeled by $A_{\mu}^{\Lambda}\ (\Lambda=0,1,\cdots, n_v)$.

\subsubsection*{Vector sector} 
The vector sector is governed by the prepotential $F(X^{\Lambda})$, which is a holomorphic function of $n_v+1$ complex variables $X^{\Lambda}\ (\Lambda=0,1,\cdots, n_v)$, and is homogeneous of degree two. In general, it can be parameterized as 
\begin{align}
F=-i(X^0)^2f(X^{i}/X^0),
\end{align}
where $f$ is an arbitrary holomorphic function. In $\mathcal{N}=2$ supergravity, the theory has (on-shell) Sp$(2n_v+2,\mathbb{R})$ symmetry which acts on the holomorphic section,
\begin{align}
\Omega ^M(z)=\left( \begin{array}{cc} X^{\Lambda }(z)\\ F_{\Sigma }(z)\\ \end{array} \right) , \ \ (\Lambda ,\Sigma =0,1,\cdots ,n_v)
\end{align}
where $F_{\Sigma}=\partial F/\partial X^{\Sigma}$, and the index $M$ specifies $2n_v+2$ components of the symplectic vector.

Based on the holomorphic section $\Omega$, the K\"ahler potential $\mathcal{K}$ is given in a manifestly symplectic invariant way as
\begin{align}
\mathcal{K}=-\log (i \bar{\Omega}^T\mathbb{C}\Omega)=-\log \left(i  \bar{X}^{\Lambda }F_{\Lambda }-i\bar{F}_{\Lambda }X^{\Lambda }\right),
\end{align}
where $\mathbb{C}$ is a symplectic invariant tensor, 
\begin{align}
\mathbb{C}=\begin{pmatrix} 
{\bm{0}}_{n_v+1} & {\bm{1}}_{n_v+1} \\ -{\bm{1}}_{n_v+1}& {\bm{0}}_{n_v+1}\\ 
\end{pmatrix}.
\end{align}

We take a special coordinate as
\begin{align}
X^0=1, \ \ X^i=z^i,
\end{align}
where $z^i $ are identified as physical scalars in vector multiplets. Then, $F_{\Lambda}=\{F_0,F_i\}$ becomes
\begin{align}
F_0=-2if+iz^if_i,\ \ F_i=-if_i,
\end{align}
where the subscript $i$ on $f$ denotes the derivative with respect to $z^i$. The K\"ahler potential is then written by
\begin{align}
\mathcal{K}=-\log \mathcal{K}_0,\ \ {\rm{where}} \ \ \mathcal{K}_0\equiv 2(f+\bar{f})-(z-\bar{z})^i(f_i-\bar{f}_i).
\end{align}
This is the general form of the K\"ahler potential.

The derivatives of the K\"ahler potential, the K\"ahler metric, and the Levi-Civita connection are computed as 
\begin{align}
&\partial_i\mathcal{K}=-\frac{\partial_i\mathcal{K}_0}{\mathcal{K}_0}=-\frac{1}{\mathcal{K}_0}\left(  f_i+\bar{f}_i-(z-\bar{z})^jf_{ij}\right) ,\label{der_K}\\
&g_{i\bar{j}}=\partial_i\mathcal{K}\partial_{\bar{j}}\mathcal{K}-\frac{1}{\mathcal{K}_0}( f_{ij}+\bar{f}_{ij}),\label{metric_K}\\
&\Gamma ^k_{ij}=g^{k\bar{\ell}}\partial_i g_{j\bar{\ell}}=2\delta^k_{(i}\partial_{j )}\mathcal{K}-\frac{g^{k\bar{\ell}}}{\mathcal{K}_0}(\partial _i\partial _j\mathcal{K}_0\partial _{\bar{\ell}}\mathcal{K}+\partial _i\partial _j\partial _{\bar{\ell}}\mathcal{K}_0).
\end{align}
Furthermore, for later convenience, we list several quantities which appear in the Lagrangian:
\begin{align}
&V^M=\left( \begin{array}{cc} L^{\Lambda }\\ M_{\Sigma }\\ \end{array} \right) \equiv e^{\mathcal{K}/2}\Omega^M = e^{\mathcal{K}/2}\left( \begin{array}{cc} X^{\Lambda }(z)\\ F_{\Sigma }(z)\\ \end{array} \right) ,\\
&U^M_i=\left( \begin{array}{cc} f^{\Lambda }_i\\ h_{\Sigma i}\\ \end{array} \right) \equiv \mathcal{D}_i V^M=\left(  \partial_i +\frac{1}{2}\partial_i\mathcal{K}\right) V^M,\\
&\mathcal{D}_i U_j^M=\partial_iU_j^M +\frac{1}{2}\partial_i\mathcal{K} U_j^M- \Gamma ^k_{ij}U_k^M= e^{\mathcal{K}}f_{ijk}g^{k\bar{k}}\bar{U}_{\bar{k}}^M. \label{formula_1}
\end{align}
Finally, the gauge kinetic functions $\mathcal{N}_{\Lambda \Sigma}$ are given by  
\begin{align}
\mathcal{N}_{\Lambda \Sigma}=\bar{F}_{\Lambda \Sigma}+\frac{2i{\rm{Im}}F_{\Lambda \Gamma}{\rm{Im}}F_{\Sigma \Pi}X^{\Pi}X^{\Gamma}}{{\rm{Im}}F_{\Pi \Gamma}X^{\Pi}X^{\Gamma}}.\label{def_GK}
\end{align}

\subsubsection*{Hyper sector} 
As for the hyper sector, we consider the following quaternion-K\"ahler metric~\cite{Ferrara:1995gu},
\begin{align}
h_{uv}=\frac{1}{2(b^0)^2}\delta_{uv},\label{Q_metric}
\end{align}
which describes a nonlinear sigma model on SO$(4,1)/{\rm{SO}}(4)$. As shown recently in Ref.~\cite{Antoniadis:2018blk}, this is the unique metric of a single hypermultiplet for the partial breaking in Minkowski space. The vielbein $\mathcal{U}^{\alpha A}=\mathcal{U}^{\alpha A}_udb^u$ can be read off as
\begin{align}
\mathcal{U}^{\alpha A}=\frac{1}{2b^0}\epsilon^{\alpha \beta}\left(db^0-i\sum_{x=1}^3\sigma ^x d b^x\right)_{\beta}^{\ \ A}, \label{vielbein}
\end{align}
where $A =1,2$ and $\alpha =1,2$ represent the SU(2) and Sp(2) indices respectively (their conventions are shown in Appendix~\ref{notation}). $\sigma^x$ is the standard Pauli matrices.
The SU(2) connection is given by
\begin{align}
\omega_u^x=\frac{1}{b^0}\delta_u^x.
\end{align}

Note that Eq.~$\eqref{Q_metric}$ admits three commuting isometries:
\begin{align}
b^m\rightarrow b^m+c^m,\ \ (m=1,2,3)\label{shift}
\end{align}
where $c^m$ are real constants.\footnote{Although both indices $m$ and $x$ run over $1,2$ and $3$, $m$ labels the isometries while $x$ is used to emphasize the SU(2) structure.} Then, the Killing vectors $k_m^u$ which generate these transformations and the moment maps corresponding to $k_m^u$ are 
\begin{align}
k_m^u=\delta_m^u,\ \ \mathcal{P}_m^x=\frac{1}{b^0}\delta_m^x. \label{killingmap}
\end{align}

\subsection{Gauging} 
In order to discuss the supersymmetry breaking, we will gauge some of the isometries of the hyper sector. For this purpose, we employ the embedding tensor formalism~\cite{deWit:2002vt, deWit:2005ub}, 
which is useful for discussing the general gauging of the extended supergravity (see also~\cite{Trigiante:2016mnt,Samtleben:2008pe} for a review). This formalism formally introduces a double copy of the gauge fields, i.e., the electric gauge fields~$A_\mu^\Lambda$ 
and the magnetic gauge fields~$A_{\mu\Sigma}$ $(\Lambda,\Sigma=0,1,\cdots,n_v)$, 
and gauges some of the global symmetries with the gauge couplings, 
\begin{align}
\Theta_{M}^{\ m}=\left( \begin{array}{cc} \Theta_{\Lambda }^{\ m}\\ \Theta^{\Sigma m}\\ \end{array} \right), \label{dET}
\end{align}
which is called the embedding tensor.

The tensor $\Theta_{M}^{\ m}$ must satisfy several conditions for the self-consistency of the theory~\cite{deWit:2002vt, deWit:2005ub}. In our case where no isometry on the vector sector is gauged, the only corresponding constraint is 
\begin{align}
\Theta_{M}^{\ m}\mathbb{C}^{MN}\Theta_{N}^{\ n}=0.\label{cond_theta}
\end{align}
Then the covariant derivative is defined by
\begin{align}
D_{\mu}\equiv&\;
\partial_{\mu}-A_{\mu}^{\Lambda}\Theta_{\Lambda}^{\ m}T_m-A_{\mu\Sigma}\Theta^{\Sigma m}T_m, \label{def_covder}
\end{align}
where $T_m$ are generators of the isometries $\eqref{shift}$, thus $k_m^u=T_mb^u$. 
We also define 
\begin{align}
k_M^u=\Theta_{M}^{\ m}k_m^u,\ \ \mathcal{P}_M^x=\Theta_{M}^{\ m}\mathcal{P}_m^x. \label{KillingMap}
\end{align}
Note that the magnetic vectors $A_{\mu\Lambda}$ also participate in the gauging.

Our interest is the $\mathcal{N}=2$ supergravity system with two breaking scales, which can realize the $\mathcal{N}=1$ (partially broken) vacuum in some limits of the gauge couplings. 
We need to gauge two of the isometries in Eq.~$\eqref{shift}$ to obtain the partial breaking~\cite{Ferrara:1995gu,Ferrara:1995xi,Fre:1996js} because the $\mathcal{N}=1$ massive gravitino multiplet contains two massive vector fields, which come from the gauging of two isometries. Also, we need a magnetic entry in Eq.~$\eqref{dET}$ for the partial breaking 
because it is proven in Ref.~\cite{Louis:2009xd} that  purely electric gauging cannot preserve $\mathcal{N}=1$ supersymmetry. 
From these observations, we take the embedding tensor as
\begin{align}
\Theta_{M}^{\ m}=\begin{pmatrix} 
0& e_2 & e_3 \\ 0& E_i& 0\\ 0& 0&0 \\ 0& M^i&0
\end{pmatrix}, \label{theta_2}
\end{align}
where we choose the directions $b^2$ and $b^3$ to be gauged.
All of the parameters in Eq.~$\eqref{theta_2}$ are real constants. The gauge coupling constants $e_2,e_3$ and $E_i$ are the {\it electric} ones, and $M^i$ are the {\it magnetic} ones. 
One can check that Eq.~$\eqref{theta_2}$ satisfies Eq.~$\eqref{cond_theta}$. This is a generalization of the one 
discussed in Ref.~\cite{Antoniadis:2018blk}. 
Our setup reduces to the model of Ref.~\cite{Ferrara:1995gu} when $e_2=E_i=0$ and $n_v=1$.

\subsection{Action and supersymmetry transformation} \label{AST}
Here we show the relevant parts of the Lagrangian~\cite{Andrianopoli:1996cm,DallAgata:2003sjo,DAuria:2004yjt,Andrianopoli:2011zj}. Due to the existence of the embedding tensor as well as some magnetic vector fields, we have to introduce auxiliary two-form fields $B_{\mu\nu, m}$ for consistency (see Refs.~\cite{deWit:2002vt,deWit:2005ub} for detail). 

The Lagrangian is given by 
\begin{align}
\mathcal{L}=-\frac{1}{2}R+\mathcal{L}_{\rm{kin}}+\mathcal{L}_{\rm{Y}}+\mathcal{L}_{\rm{Pauli}}+\mathcal{L}_{\rm{der}}+\mathcal{L}_{\rm{top}}-V, \label{full_L}
\end{align}
where $R$ is the Ricci scalar and we have omitted the four fermi interactions. $\mathcal{L}_{\rm{kin}}$ consists of the kinetic terms and we further decompose it for later convenience,
\begin{align}
&\mathcal{L}_{\rm{kin}}=\mathcal{L}_{{\rm{kin}},z}+\mathcal{L}_{{\rm{kin}},b}+\mathcal{L}_{{\rm{kin}},v}+\mathcal{L}_{{\rm{kin}},f}\\
&\mathcal{L}_{{\rm{kin}},z}=g_{i\bar{j}}\partial _{\mu}z^i\partial ^{\mu}\bar{z}^{\bar{j}} \label{kin,z}\\
&\mathcal{L}_{{\rm{kin}},b}=h_{uv}D _{\mu}b^uD ^{\mu}b^v \label{kin,b}\\
&\mathcal{L}_{{\rm{kin}},v}=\frac{1}{4}\mathcal{I}_{\Lambda \Sigma}\mathcal{H}_{\mu \nu }^{\Lambda }\mathcal{H}^{\Sigma \mu \nu }+\frac{i}{4}\mathcal{R}_{\Lambda \Sigma}\mathcal{H}_{\mu\nu}^{\Lambda }\tilde{\mathcal{H}}^{\Sigma \mu\nu},\label{kin,v}\\
&\mathcal{L}_{{\rm{kin}},f}=\frac{1}{\sqrt{-g}}\varepsilon^{\mu\nu\rho\sigma}\bar{\psi}_{\mu}^A\gamma_{\nu}\mathcal{D}_{\rho}\psi_{A\sigma}-\frac{i}{2}  g_{i\bar{j}}\bar{\lambda}^{iA}\gamma^{\mu}\mathcal{D}_{\mu}\lambda_A^{\bar{j}}-i\bar{\zeta}^{\alpha}\gamma^{\mu}\mathcal{D}_{\mu}\zeta_{\alpha}+{\rm{h.c.}},\label{kin,f}
\end{align}
where $\mathcal{H}_{\mu \nu}^{\Lambda}$ is a gauge invariant combination of the field-strength and the two-form field,
\begin{align}
\mathcal{H}_{\mu \nu}^{\Lambda} \equiv F_{\mu \nu}^{\Lambda}+\frac{1}{2}\Theta ^{\Lambda m}B_{\mu\nu ,m}, \label{def_H}
\end{align}
and $\tilde{\mathcal{H}}_{\mu\nu}\equiv -\frac{i}{2}\varepsilon_{\mu\nu\rho\sigma}\mathcal{H}^{\rho\sigma}$. $\mathcal{I}_{\Lambda \Sigma}$ and $\mathcal{R}_{\Lambda \Sigma}$ are imaginary and real parts of $\mathcal{N}_{\Lambda \Sigma}$. $D_{\mu}b^u$ is defined in Eq.~$\eqref{def_covder}$.
The covariant derivatives for fermions are shown in Appendix~\ref{int}.

$\mathcal{L}_{\rm{Y}}$ denotes the interactions including the fermion bilinears and it is explicitly given by
\begin{align}
\nonumber \mathcal{L}_{\rm{Y}}=&2\mathbb{S}_{AB}\bar{\psi }^A_{\mu}\gamma ^{\mu\nu}\psi _{\nu}^B +ig_{i\bar{j}}W^{iAB}\bar{\lambda }_A^{\bar{j}}\gamma _{\mu}\psi ^{\mu}_B+2iN_{\alpha }^A\bar{\zeta }^{\alpha }\gamma _{\mu}\psi ^{\mu}_A \\
&+\mathcal{M}^{\alpha \beta }\bar{\zeta }_{\alpha }\zeta _{\beta }+\mathcal{M}^{\alpha}_{i  B }\bar{\zeta }_{\alpha }\lambda ^{iB}+\mathcal{M}_{ijAB}\bar{\lambda }^{iA}\lambda ^{jB}+{\rm{h.c.}}, \label{F_mass}
\end{align}
where
\begin{align}
&\mathbb{S}_{AB}\equiv \frac{i}{2}(\sigma _x)_{AB}\mathcal{P}_M^xV^M,\label{psipsi}\\
&W^{iAB}\equiv i(\sigma _x)^{AB}\mathcal{P}_M^xg^{i\bar{j}}\bar{U}_{\bar{j}}^M,\\
&N_{\alpha }^A\equiv -2\mathcal{U}^A_{u \alpha }k_M^u\bar{V}^M,\\
&\mathcal{M}^{\alpha \beta }\equiv \mathcal{U}^{\alpha A}_{ u}\mathcal{U}^{\beta B}_{ v}\epsilon _{AB}\mathcal{D} ^{[u}k_M^{v]}V^M,\\
&\mathcal{M}^{\alpha}_{i B}\equiv 4\mathcal{U}^{\alpha }_{B u}k_M^uU_i^M,\\
&\mathcal{M}_{ABij}\equiv  \frac{i}{2}(\sigma _x)_{AB}\mathcal{P}_M^x\mathcal{D}_iU_j^M. \label{zz}
\end{align}
In their expressions, $\mathcal{D}_uk_M^v=\partial_uk_M^v+\epsilon^{vxy}\omega_u^xk_M^y$ and $\mathcal{D}_iU_j^M$ can be found in Eq.~$\eqref{formula_1}$.  

The interaction terms $\mathcal{L}_{\rm{Pauli}}$ and $\mathcal{L}_{\rm{der}}$ are necessary in subsection~\ref{summary} and their explicit expressions are shown in Appendix~\ref{int}.

$\mathcal{L}_{\rm{top}}$ is required for consistency of the embedding tensor formalism and it is given by~\cite{deWit:2005ub} 
\begin{align}
\mathcal{L}_{\rm{top}}=-\frac{i}{4}\Theta^{\Lambda m}\tilde{B}_{\mu \nu ,m}\left( F_{\Lambda}^{ \mu \nu}-\frac{1}{4}\Theta _{\Lambda}^{\ n}B^{\mu \nu }_n\right). \label{L_top}
\end{align}

Finally, the scalar potential $V$ is given by
\begin{align}
V=g^{i\bar{j}}U_i^M\bar{U}_{\bar{j}}^N\mathcal{P}_M^x\mathcal{P}_N^x+4h_{uv}k^u_Mk^v_N\bar{V}^MV^N-3\bar{V}^MV^N\mathcal{P}_M^x\mathcal{P}_N^x.\label{SP_general}
\end{align}

Note that the Lagrangian $\eqref{full_L}$ has larger gauge symmetry, in addition to the Abelian gauge symmetry of the vector fields. Indeed, it is invariant under 
\begin{align}
&\delta A_{\mu}^{\Lambda}=-\frac{1}{2}\Theta^{\Lambda m}\Xi_{\mu m},\\
&\delta A_{\mu\Lambda}=\frac{1}{2}\Theta_{\Lambda}^{\ m}\Xi_{\mu m},\\
&\delta B_{\mu\nu ,m}=2\partial_{[\mu}\Xi_{\nu ] m},
\end{align}
where $\Xi_{\mu m}$ are the gauge transformation parameters of the two-forms.

We need the supersymmetry transformations of the fermions to discuss supersymmetry breaking. The relevant parts are given by
\begin{align}
&\delta \psi_{A\mu}=i\mathbb{S}_{AB}\gamma_{\mu}\epsilon^B+\cdots,\label{T_psi}\\
&\delta \lambda^{iA}=W^{iAB}\epsilon_B+\cdots,\label{T_lambda}\\
&\delta \zeta_{\alpha}=N_{\alpha}^A\epsilon_A+\cdots,\label{T_zeta}
\end{align}
where $\epsilon^A$ are the transformation parameters. The ellipses represent other contributions, which vanish in the Minkowski vacuum.

\section{Spectrum}\label{Spectrum}
Here we derive the mass spectrum of our model.

\subsection{Scalar potential and minimization} \label{Minimization}
First, let us discuss conditions the vacuum satisfies. Under the gauging~$\eqref{theta_2}$, the scalar potential~$\eqref{SP_general}$ is explicitly given by 
\begin{align}
V=\frac{e^{\mathcal{K}}}{(b^0)^2}\biggl[g^{i\bar{j}}D^x_i\bar{D}^x_{\bar{j}}-\left( \mathcal{E}^x-if_j\mathcal{M}^{xj}+\mathcal{N}_j^xz^j \right)\left( \mathcal{E}^x+i\bar{f}_k\mathcal{M}^{xk}+\mathcal{N}_k^x\bar{z}^k \right)\biggr] , \label{SV}
\end{align}
where we have defined
\begin{align}
&D^x_i\equiv \mathcal{E}^x\partial_i\mathcal{K}-i\mathcal{M}^{xj} (f_{ij}+\partial_i\mathcal{K}f_j)+\mathcal{N}_j^x(\delta_i^j+\partial_i\mathcal{K}z^j)  ,\\ 
&\mathcal{E}^x\equiv (0,e_2,e_3), \ \ \mathcal{M}^{xj} \equiv (0,M^j,0),\ \ \mathcal{N}_j^x\equiv (0,E_j,0).
\end{align}
Note that the scalar potential $\eqref{SV}$ is independent of $b^m\ (m=1,2,3)$.
Thus, the minimum is obtained by solving 
\begin{align}
&\partial_i V=0,\label{cond_0}\\
&\partial_{b^0} V=0, \label{cond_2}
\end{align}
where $\partial_i=\partial /\partial z^i$ and $\partial_{b^0}=\partial /\partial b^0$. They give $n_v+1$ equations in general. By using Eq.~$\eqref{formula_1}$, Eq.~$\eqref{cond_0}$ can be summarized as
\begin{align}
f_{ijk}g^{j\bar{j}}g^{k\bar{k}}\bar{D}^x_{\bar{j}}\bar{D}^x_{\bar{k}}=0.\label{cond_1}
\end{align}
In general, it is difficult to solve these equations for a general form of $f$. Therefore, for simplicity, we assume that one of $z^i$ has a nonzero expectation value (we choose it the $N$-th direction), and the vacuum satisfies the following conditions:
\begin{align}
&\left< z^i\right>=\delta^{iN}\lambda,\label{z_assum}\\
&\left< f_i\right>=\delta_{iN}\left< f_N\right>,\\
&\left< f_{iN}\right>=\delta_{iN}\left< f_{NN}\right>,\\
&\left< f_{NNi}\right>=\delta_{iN}\left< f_{NNN}\right>.\label{f'''_assum}
\end{align}
Then, the derivative of the K\"ahler potential and the K\"ahler metric become  
\begin{align}
&\left< \partial _i \mathcal{K}\right>=\delta_{iN}\left< \partial _N \mathcal{K}\right>,\\
& \left< g_{i\bar{N}}\right>=\delta_{i\bar{N}}\left(|\left< \partial _N \mathcal{K}\right>|^2-\left< f_{NN}+\bar{f}_{NN}\right>/\left<\mathcal{K}_0\right> \right).
\end{align}
Later, we will check that these assumptions for the vacuum are valid in a concrete choice of $f$. Furthermore, we assume that the gauging is done only for the $N$-th direction, i.e.,
\begin{align}
\mathcal{M}^{xj}=\delta^{jN}M^N, \ \ \mathcal{N}^{x}_j=\delta_{jN}E_N, \label{MN}
\end{align}
which lead to 
\begin{align}
\left< D^x_i\right>=\delta_{iN}\left< D^x_N\right>.
\end{align}

Under these simplifications, Eq.~$\eqref{cond_1}$ is reduced to just one equation,
\begin{align}
\left<f_{NNN}(g^{N\bar{N}})^2\bar{D}^x_{\bar{N}}\bar{D}^x_{\bar{N}}\right>=0,\label{cond_1'}
\end{align}
and the others are trivially satisfied.
Therefore, if the vacuum satisfies either 
\begin{align}
\left<(\bar{D}^x_{\bar{N}})^2\right>=0, \label{cond_1''}
\end{align}
or
\begin{align}
\left<f_{NNN}\right>=0, \label{cond_1'''}
\end{align}
then Eq.~$\eqref{cond_0}$ is satisfied at that point.

Next, one can check that Eq.~$\eqref{cond_2}$ is satisfied if 
\begin{align}
\left< f_{NN}\right>=-i\frac{E_N}{M^N}. \label{vev2}
\end{align}
In this case, we obtain $\left<\partial_N K\right>=\left<\partial_{\bar{N}} K\right>$ and $\left<g_{N\bar{N}}\right>=|\left< \partial_{\bar{N}} K\right>|^2$ since $\left< f_{NN}\right>$ is pure imaginary (see Eqs.~$\eqref{der_K}$ and $\eqref{metric_K}$). Also, the condition $\eqref{vev2}$ leads to the Minkowski vacuum.

Under the condition~$\eqref{vev2}$, Eq.~$\eqref{cond_1''}$ can be simplified further as,
\begin{align}
\left< f_N\right>=-\frac{1}{M^N}\left(ie_2+iE_N\lambda \pm e_3\right). \label{vev1}
\end{align}
In the following discussion, we take the plus branch.\footnote{For the choice of the minus sign, see the comment below Eq.~$\eqref{(i)Tzeta}$.}

In summary, under the assumptions~$\eqref{z_assum}$-$\eqref{f'''_assum}$ and $\eqref{MN}$, the vacuum must satisfy either of the two conditions:
\begin{enumerate}
\renewcommand{\labelenumi}{(\roman{enumi})}
\item Eqs.~$\eqref{vev2}$ and~$\eqref{cond_1'''}$,
\item Eqs.~$\eqref{vev2}$ and~$\eqref{vev1}$.
\end{enumerate}
These conditions have been derived in Ref.~\cite{Antoniadis:2018blk} in the case of a single vector multiplet ($n_v=1$). We will investigate how many supersymmetries are (un)broken in the vacuum in the next subsection.

\subsection{Supersymmetry transformation}
Let us begin with the case (ii). Under the conditions $\eqref{vev2}$ and $\eqref{vev1}$, the supersymmetry transformations~$\eqref{T_psi}$-$\eqref{T_zeta}$ in the vacuum become 
\begin{align}
&\left<  \delta \psi _{A\mu}\right> =\left<  \frac{e_3e^{\mathcal{K}/2}}{2b^0}\right> \left(\begin{array}{cc}  -1&1 \\ 1& -1\\ \end{array} \right)  \gamma _{\mu} \left( \begin{array}{cc} \epsilon ^1\\ \epsilon ^2\\ \end{array} \right) ,\label{Tpsi_1}\\
&\left<  \delta \lambda ^{iA}\right> =\delta^{iN}\left<\frac{ie_3e^{\mathcal{K}/2}}{b^0}g^{N\bar{N}}\partial _{\bar{N}}\mathcal{K}\right> \left(\begin{array}{cc}  -1&1 \\ 1& -1\\ \end{array} \right)\left( \begin{array}{cc} \epsilon _1\\ \epsilon _2\\ \end{array} \right) ,\\
&\left<  \delta \zeta _{\alpha }\right> =\left<  \frac{ie_3e^{\mathcal{K}/2}}{b^0}\right>\left(\begin{array}{cc}  -1& 1 \\  -1& 1\\ \end{array} \right) \left( \begin{array}{cc} \epsilon _1\\ \epsilon _2\\ \end{array} \right) .
\end{align}
As can be seen, all of the matrices have a zero eigenvalue. Indeed, defining 
\begin{align}
\phi_{\pm}=\frac{1}{\sqrt{2}}(\phi_1\pm \phi_2), \label{newbase}   
\end{align}
with $\phi=\{\psi,\lambda,\zeta,\epsilon \}$, we can rewrite them as  
\begin{align}
&\left( \begin{array}{cc} \left<  \delta \psi _{+\mu}\right>\\ \left<  \delta \psi _{-\mu}\right>\\ \end{array} \right)=\left< -  \frac{e_3e^{\mathcal{K}/2}}{b^0}\right>   \gamma _{\mu} \left( \begin{array}{cc} 0\\ \epsilon ^-\\ \end{array} \right) ,\\
&\left( \begin{array}{cc} \left<  \delta \lambda ^{i+}\right>\\ \left<  \delta \lambda ^{i-}\right>\\ \end{array} \right) =\delta^{iN} \left<-\frac{2ie_3e^{\mathcal{K}/2}}{b^0}g^{N\bar{N}}\partial _{\bar{N}}\mathcal{K}\right> \left( \begin{array}{cc} 0\\ \epsilon _-\\ \end{array} \right),\\
&\left( \begin{array}{cc} \left<  \delta \zeta _{+}\right>\\ \left<  \delta \zeta _{- }\right>\\ \end{array} \right) =\left<  -\frac{2ie_3e^{\mathcal{K}/2}}{b^0}\right>\left( \begin{array}{cc} \epsilon _-\\ 0\\ \end{array} \right) . \label{(i)Tzeta}
\end{align}
Therefore, one of the two supersymmetries is broken (for $\epsilon _-$ direction), but the other one is still preserved (for $\epsilon _+$ direction). The minus sign in Eq.~$\eqref{vev1}$ leads to the preservation of $\epsilon _-$ direction. We conclude that for nonzero $e_3$ and $M^N$, we always have $\mathcal{N}=1$ preserving vacuum in the case (ii).\footnote{The global supersymmetry limit of this partial breaking vacuum have been discussed in Refs.~\cite{Ferrara:1995xi,David:2003dh, Andrianopoli:2015rpa,Laamara:2017hdl}.}  

In the case (i), on the other hand, the supersymmetry transformations are
\begin{align}
&\left<  \delta \psi _{A\mu}\right> =\left<  \frac{e^{\mathcal{K}/2}}{2b^0}\right> \left(\begin{array}{cc} \tau &e_3\\ e_3& \tau\\ \end{array} \right)  \gamma _{\mu} \left( \begin{array}{cc} \epsilon ^1\\ \epsilon ^2\\ \end{array} \right) ,\label{psiM}\\
&\left<  \delta \lambda ^{iA}\right> =\delta^{iN}\left<\frac{ie^{\mathcal{K}/2}}{b^0}g^{N\bar{N}}\partial _{\bar{N}}\mathcal{K}\right> \left(\begin{array}{cc} \bar{\tau}&e_3 \\ e_3& \bar{\tau}\\ \end{array} \right)\left( \begin{array}{cc} \epsilon _1\\ \epsilon _2\\ \end{array} \right) ,\\
&\left<  \delta \zeta _{\alpha }\right> =\left<-  \frac{ie^{\mathcal{K}/2}}{b^0}\right>\left(\begin{array}{cc}  e_3&\bar{\tau} \\ -\bar{\tau}& -e_3\\ \end{array} \right) \left( \begin{array}{cc} \epsilon _1\\ \epsilon _2\\ \end{array} \right) ,\label{zetaM}
\end{align}
where
\begin{align}
\tau \equiv ie_2+iE_N\lambda+M^N\left<  f_N\right>.
\end{align}
In this case, the value of $\left<  f_N\right>$ is not determined by the vacuum conditions. The vacuum can be broken to $\mathcal{N}=0$ depending on $\left<  f_N\right>$. 
If we choose $\left< f_N\right>$ as Eq.(\ref{vev1}), the situation goes back to the case~(ii), and ${\cal N}=1$ supersymmetry 
is preserved. Thus, we can interpolate the full and the partial breakings in this case.\footnote{In the purely electric gauging, it is possible to break full $\mathcal{N}=2$ supersymmetry, but the supplemental condition for the partial breaking~$\eqref{vev1}$ is never satisfied~\cite{Louis:2009xd}.}

In order to characterize the deviation from the partial-breaking condition~$\eqref{vev1}$, we parametrize $\left<  f_N\right>$ as 
\begin{align}
\left< f_N\right>=-\frac{1}{M^N}\left(ie_2+iE_N\lambda + e_3+\Delta \right),
\end{align}
where we have introduced $\Delta$ (complex) as an order parameter of $\mathcal{N}=0$ breaking. $\Delta=0$ recovers $\mathcal{N}=1$ preserving (partially broken) vacuum. Then, Eqs.~$\eqref{psiM}$-$\eqref{zetaM}$ are expressed as
\begin{align}
&\left<  \delta \psi _{A\mu}\right> =\left< - \frac{e^{\mathcal{K}/2}}{2b^0}\right> \left(\begin{array}{cc}  e_3+\Delta &-e_3 \\ -e_3&  e_3+\Delta\\ \end{array} \right)  \gamma _{\mu} \left( \begin{array}{cc} \epsilon ^1\\ \epsilon ^2\\ \end{array} \right) ,\\
&\left<  \delta \lambda ^{iA}\right> =\delta^{iN}\left<-\frac{ie^{\mathcal{K}/2}}{b^0}g^{N\bar{N}}\partial _{\bar{N}}\mathcal{K}\right> \left(\begin{array}{cc}  e_3+\bar{\Delta}&-e_3 \\ -e_3&  e_3+\bar{\Delta}\\ \end{array} \right)\left( \begin{array}{cc} \epsilon _1\\ \epsilon _2\\ \end{array} \right) ,\\
&\left<  \delta \zeta _{\alpha }\right> =\left<-  \frac{ie^{\mathcal{K}/2}}{b^0}\right>\left(\begin{array}{cc}  e_3&- e_3-\bar{\Delta} \\   e_3+\bar{\Delta}& -e_3\\ \end{array} \right) \left( \begin{array}{cc} \epsilon _1\\ \epsilon _2\\ \end{array} \right) .
\end{align}
In terms of the basis $\eqref{newbase}$, these are rewritten as 
\begin{align}
&\left( \begin{array}{cc} \left<  \delta \psi _{+\mu}\right>\\ \left<  \delta \psi _{-\mu}\right>\\ \end{array} \right)=\left<-  \frac{e^{\mathcal{K}/2}}{2b^0}\right>   \gamma _{\mu} \left( \begin{array}{cc} \Delta \epsilon ^+ \\ (2e_3+\Delta)\epsilon ^-\\ \end{array} \right) ,\label{psipm}\\
&\left( \begin{array}{cc} \left<  \delta \lambda ^{i+}\right>\\ \left<  \delta \lambda ^{i-}\right>\\ \end{array} \right) =\delta^{iN} \left<-\frac{ie^{\mathcal{K}/2}}{b^0}g^{N\bar{N}}\partial _{\bar{N}}\mathcal{K}\right> \left( \begin{array}{cc} \bar{\Delta} \epsilon _+\\ (2e_3+\bar{\Delta})\epsilon _-\\ \end{array} \right),\\
&\left( \begin{array}{cc} \left<  \delta \zeta _{+}\right>\\ \left<  \delta \zeta _{- }\right>\\ \end{array} \right) =\left<-  \frac{ie^{\mathcal{K}/2}}{b^0}\right>\left( \begin{array}{cc} (2e_3+\bar{\Delta})\epsilon _- \\ -\bar{\Delta}\epsilon _+\\ \end{array} \right) .\label{zetapm}
\end{align}
The full supersymmetry breaking occurs unless $\Delta=0,-2e_3$. 
In the following, we call the supersymmetry transformations caused by $\epsilon_+$ and $\epsilon_-$ as the first and the second supersymmetry transformations, respectively.

\subsection{Explicit model}\label{Explicit}
Let us specify our model. We assume that $f$ has the following form,\footnote{In Ref.~\cite{Ferrara:1995gu}, the case $f=z$ is considered.}
\begin{align}
f=c_0+c_iz^i+\frac{1}{2}c_{ij}z^iz^j+\frac{1}{6}c_{ijk}z^iz^jz^k, \label{f=}
\end{align}
where $c_0, c_i, c_{ij}$ and $c_{ijk}$ are complex constants and totally symmetric for their indices. This satisfies Eqs.~$\eqref{cond_1'''}$ and $\eqref{vev2}$ if
\begin{align}
&c_{NNN}=0,\\
&c_{NN}=-i\frac{E_N}{2M^N}.
\end{align}
Also, from the assumptions~$\eqref{z_assum}$-$\eqref{f'''_assum}$, we obtain
\begin{align}
f=c_0+c_Nz^N+c_{NN}(z^N)^2+c_{\hat{a}\hat{b}}z^{\hat{a}}z^{\hat{b}}+3c_{\hat{a}\hat{b}N}z^{\hat{a}}z^{\hat{b}}z^N+c_{\hat{a}\hat{b}\hat{c}}z^{\hat{a}}z^{\hat{b}}z^{\hat{c}},\label{example_f}
\end{align}
where we have divided the indices $i=\{\hat{a}, N\}$ with $\hat{a}=1,\cdots, n_v-1$.\footnote{The convention of the indices is summarized in Appendix~\ref{notation}. }

\subsection{Fermion mass} \label{Fermionmass}
Let us check the spectrum of the fermion sector. In the vacuum, fermion mass terms from the interactions in Eq.~$\eqref{F_mass}$ are evaluated as 
\begin{align}
\nonumber \mathcal{L}_{\rm{Y}}=&\left<  \frac{ie^{\mathcal{K}/2}}{b^0}\right> \left(\Delta\psi _{\mu}^+\gamma ^{\mu \nu }\psi _{\nu}^++(\Delta+2e_3)\psi _{\mu}^-\gamma ^{\mu \nu }\psi _{\nu}^-\right)\\
\nonumber &+\left<  \frac{e^{\mathcal{K}/2}}{b^0}\partial_{\bar{N}}\mathcal{K}\right> \left(\bar{\Delta}\bar{\lambda }_+^{\bar{N}}\gamma _{\mu}\psi ^{\mu}_++(\bar{\Delta}+2e_3)\bar{\lambda }_-^{\bar{N}}\gamma _{\mu}\psi ^{\mu}_-\right)\\
\nonumber &+\left<  \frac{2e^{\mathcal{K}/2}}{b^0}\right>\left((\bar{\Delta}+2e_3)\bar{\zeta }^+\gamma _{\mu}\psi ^{\mu}_--\bar{\Delta}\bar{\zeta }^-\gamma _{\mu}\psi ^{\mu}_+\right)\\
\nonumber &+\left<  \frac{ie^{\mathcal{K}/2}}{b^0}\right> \left((\Delta +2e_3)\bar{\zeta }_+\zeta _+ +\Delta\bar{\zeta }_-\zeta _-\right)\\
\nonumber &+\left<  \frac{2ie^{\mathcal{K}/2}}{b^0}\partial_N\mathcal{K}\right> \left((\Delta+2e_3)\bar{\zeta }_+\lambda ^{N-}-\Delta\bar{\zeta }_-\lambda ^{N+}\right)\\
\nonumber  &-\left<  \frac{ie^{3\mathcal{K}/2}}{2b^0}g^{N\bar{N}}\partial_{\bar{N}}\mathcal{K}f_{\hat{a}\hat{b}N}\right>\left(\bar{\Delta}\bar{\lambda}^{\hat{a}-}\lambda^{\hat{b}-}+(\bar{\Delta}+2e_3)\bar{\lambda}^{\hat{a}+}\lambda^{\hat{b}+}\right)+{\rm{h.c.}}\\
\nonumber  \equiv & m_1\psi _{\mu}^+\gamma ^{\mu \nu }\psi _{\nu}^+ +im_1\bar{\chi}_{\bullet}\gamma _{\mu}\psi^{\mu +}+\frac{m_1}{3}\bar{\chi}_{\bullet}\chi_{\bullet}-\frac{m_1}{3}\bar{\eta}_{\bullet}\eta_{\bullet}\\
\nonumber  &+ m_2\psi _{\mu}^-\gamma ^{\mu \nu }\psi _{\nu}^- +im_2\bar{\tilde{\chi}}_{\bullet}\gamma _{\mu}\psi^{\mu -}+\frac{m_2}{3}\bar{\tilde{\chi}}_{\bullet}\tilde{\chi}_{\bullet}-\frac{m_2}{3}\bar{\tilde{\eta}}_{\bullet}\tilde{\eta}_{\bullet}\\
 &+\sum_{\hat{a}=1}^{N-1}\left(\frac{1}{2}Gm_+\bar{\lambda}^{\hat{a}+}\lambda^{\hat{a}+}+\frac{1}{2}Gm_-\bar{\lambda}^{\hat{a}-}\lambda^{\hat{a}-}\right)+{\rm{h.c.}}.\label{Fmass}
\end{align}
In the second line, we have defined 
\begin{align}
&m_1=\left<  \frac{ie^{\mathcal{K}/2}}{b^0}\right>\Delta ,\ \ m_2=\left<  \frac{ie^{\mathcal{K}/2}}{b^0}\right>(\Delta+2e_3),\\
&m_-=\left<-  \frac{ie^{3\mathcal{K}/2}}{b^0}\sqrt{g^{N\bar{N}}}\frac{C}{G}\right>\bar{\Delta}, \ \ m_+=\left< - \frac{ie^{3\mathcal{K}/2}}{b^0}\sqrt{g^{N\bar{N}}}\frac{C}{G}\right>(\bar{\Delta}+2e_3),\\
&\chi_{\bullet}=\left<  \partial_{N}\mathcal{K}\right>\lambda^{N+}-2\zeta_-,\ \ \eta_{\bullet}=\left<  \partial_{N}\mathcal{K}\right>\lambda^{N+}+\zeta_-,\\
&\tilde{\chi}_{\bullet}=\left<  \partial_{N}\mathcal{K}\right>\lambda^{N-}+2\zeta_+,\ \ \tilde{\eta}_{\bullet}=\left<  \partial_{N}\mathcal{K}\right>\lambda^{N-}-\zeta_+.
\end{align}
Also, we have assumed that  
\begin{align}
\left<f_{\hat{a}\hat{b}N}\right>=C\delta_{\hat{a}\hat{b}},\ \ \left< g_{\hat{a}\bar{\hat{b}}}\right>=G\delta_{\hat{a}\bar{\hat{b}}}, \label{diag}
\end{align}
where $C$ and $G$ are complex and real constants, for simplicity.\footnote{These can be achieved by choosing $c_{\hat{a}\hat{b}}, c_{\hat{a}\hat{b}N} \propto \delta_{\hat{a}\hat{b}} $ in Eq.~$\eqref{example_f}$.}

From the kinetic terms in Eq.~$\eqref{kin,f}$, we obtain
\begin{align}
\nonumber  \mathcal{L}_{{\rm{kin}},f}=&\frac{\varepsilon^{\mu\nu\rho\sigma}}{\sqrt{-g}}\left(\bar{\psi}_{\mu}^+\gamma_{\nu}\partial_{\rho}\psi_{+\sigma}+\bar{\psi}_{\mu}^-\gamma_{\nu}\partial_{\rho}\psi_{-\sigma}\right)-\frac{i}{2}\left<  g_{N\bar{N}}\right>\left(\bar{\lambda}^{N+}\gamma^{\mu}\partial_{\mu}\lambda_{+}^{\bar{N}}+\bar{\lambda}^{N-}\gamma^{\mu}\partial_{\mu}\lambda_{-}^{\bar{N}}\right)\\
&+ \sum_{\hat{a}=1}^{N-1}\left(-\frac{i}{2}G\bar{\lambda}^{\hat{a}+}\gamma^{\mu}\partial_{\mu}\lambda_{+}^{\bar{\hat{a}}}-\frac{i}{2}G\bar{\lambda}^{\hat{a}-}\gamma^{\mu}\partial_{\mu}\lambda_{-}^{\bar{\hat{a}}}\right)-i\left(\bar{\zeta}^{+}\gamma^{\mu}\partial_{\mu}\zeta_{+}+\bar{\zeta}^{-}\gamma^{\mu}\partial_{\mu}\zeta_{-}\right)+{\rm{h.c.}}.
\end{align}
In terms of $\chi (\tilde{\chi})$ and $\eta(\tilde{\eta})$ defined above, we can rewrite them as  
\begin{align}
\nonumber  \mathcal{L}_{{\rm{kin}},f}=&\frac{\varepsilon^{\mu\nu\rho\sigma}}{\sqrt{-g}}\left(\bar{\psi}_{\mu}^+\gamma_{\nu}\partial_{\rho}\psi_{+\sigma}+\bar{\psi}_{\mu}^-\gamma_{\nu}\partial_{\rho}\psi_{-\sigma}\right)+\sum_{\hat{a}=1}^{N-1}\left(-\frac{i}{2}G\bar{\lambda}^{\hat{a}+}\gamma^{\mu}\partial_{\mu}\lambda_{+}^{\bar{\hat{a}}}-\frac{i}{2}G\bar{\lambda}^{\hat{a}-}\gamma^{\mu}\partial_{\mu}\lambda_{-}^{\bar{\hat{a}}}\right)\\
&-\frac{i}{6}\bar{\chi}_{\bullet}\gamma^{\mu}\partial_{\mu}\chi^{\bullet}-\frac{i}{3}\bar{\eta}_{\bullet}\gamma^{\mu}\partial_{\mu}\eta^{\bullet}-\frac{i}{6}\bar{\tilde{\chi}}_{\bullet}\gamma^{\mu}\partial_{\mu}\tilde{\chi}^{\bullet}-\frac{i}{3}\bar{\tilde{\eta}}_{\bullet}\gamma^{\mu}\partial_{\mu}\tilde{\eta}^{\bullet}+{\rm{h.c.}}. \label{Fkin}
\end{align}

In Eqs.~$\eqref{Fmass}$ and ~$\eqref{Fkin}$, $\chi$ and $\tilde{\chi}$ are the goldstinos which correspond to the first and the second supersymmetry breaking, respectively. It can be checked directly by 
\begin{align}
\delta \chi_{\bullet}=  \left< -  \frac{3ie^{\mathcal{K}/2}}{b^0}\right> \bar{\Delta}\epsilon_+ , \ \ \delta \tilde{\chi}_{\bullet}=  \left< -  \frac{3ie^{\mathcal{K}/2}}{b^0}\right> (2e_3+\bar{\Delta})\epsilon_- .
\end{align}
Thus, in the unitary gauge, $\chi=\tilde{\chi}=0$, they disappear from the spectrum. Then, we obtain
\begin{align}
\nonumber  \mathcal{L}_{{\rm{kin}},f}+\mathcal{L}_{\rm{Y}}=&\frac{\varepsilon^{\mu\nu\rho\sigma}}{\sqrt{-g}}\bar{\psi}_{\mu}^+\gamma_{\nu}\partial_{\rho}\psi_{+\sigma}-\frac{i}{3}\bar{\eta}_{\bullet}\gamma^{\mu}\partial_{\mu}\eta^{\bullet}+m_1\psi _{\mu}^+\gamma ^{\mu \nu }\psi _{\nu}^+ -\frac{m_1}{3}\bar{\eta}_{\bullet}\eta_{\bullet}\\
\nonumber &+\frac{\varepsilon^{\mu\nu\rho\sigma}}{\sqrt{-g}}\bar{\psi}_{\mu}^-\gamma_{\nu}\partial_{\rho}\psi_{-\sigma}-\frac{i}{3}\bar{\tilde{\eta}}_{\bullet}\gamma^{\mu}\partial_{\mu}\tilde{\eta}^{\bullet}+ m_2\psi _{\mu}^-\gamma ^{\mu \nu }\psi _{\nu}^- -\frac{m_2}{3}\bar{\tilde{\eta}}_{\bullet}\tilde{\eta}_{\bullet}\\
\nonumber &+ \sum_{\hat{a}=1}^{N-1}\left(-\frac{i}{2}G\bar{\lambda}^{\hat{a}+}\gamma^{\mu}\partial_{\mu}\lambda_{+}^{\bar{\hat{a}}}-\frac{i}{2}G\bar{\lambda}^{\hat{a}-}\gamma^{\mu}\partial_{\mu}\lambda_{-}^{\bar{\hat{a}}}\right)\\
&+\sum_{\hat{a}=1}^{N-1}\left(\frac{1}{2}Gm_+\bar{\lambda}^{\hat{a}+}\lambda^{\hat{a}+}+\frac{1}{2}Gm_-\bar{\lambda}^{\hat{a}-}\lambda^{\hat{a}-}\right)+{\rm{h.c.}}.
\end{align}
This gives the free parts of the fermion sector. 
Note that we have two massive pairs, $\psi_+, \eta$ and $\psi_-, \tilde{\eta}$, whose masses are given by $|m_1|$ and $|m_2|$ respectively. In addition, there are $2n_v-2$ massive gauginos. The $n_v-1$ gauginos ($\lambda^{\hat{a} +}$) have the mass $|m_+|$, and the others ($\lambda^{\hat{a} -}$) have $|m_-|$.\footnote{$\lambda^{\hat{a}\pm}\rightarrow \frac{1}{\sqrt{G}}\lambda^{\hat{a}\pm}$ and $\eta_{\bullet}(\tilde{\eta}_{\bullet})\rightarrow \sqrt{\frac{3}{2}}\eta_{\bullet}\left(\sqrt{\frac{3}{2}}\tilde{\eta}_{\bullet}\right)$ lead to the canonical normalization.}
Their mass scales are splitting due to the existence of $e_3$.

\subsection{Boson mass} \label{Bosonmass}
First, let us focus on the following terms
\begin{align}
\nonumber  \mathcal{L}_B\equiv &\mathcal{L}_{{\rm{kin}},v}+\mathcal{L}_{{\rm{kin}},b}+\mathcal{L}_{\rm{top}}\\
\nonumber =&\frac{1}{4}  \mathcal{I}_{\Lambda \Sigma}\mathcal{H}_{\mu \nu}^{\Lambda}\mathcal{H}^{\Sigma \mu \nu}+\frac{i}{4}  \mathcal{R}_{\Lambda \Sigma}\mathcal{H}_{\mu \nu}^{\Lambda}\tilde{\mathcal{H}}^{\Sigma \mu \nu}\\
&+h_{uv}D_{\mu}b^uD^{\mu}b^v-\frac{i}{4}\Theta^{\Lambda m}\tilde{B}_{\mu \nu ,m}\left( F_{\Lambda}^{ \mu \nu}-\frac{1}{4}\Theta _{\Lambda}^{\ n}B^{\mu \nu }_n\right). \label{LB}
\end{align}
Each term can be decomposed explicitly as  
\begin{align}
&\mathcal{H}_{\mu \nu}^{\Lambda}=F_{\mu \nu}^{\Lambda}+\frac{1}{2}\Theta^{\Lambda m}B_{\mu \nu ,m}=\left( \begin{array}{cc} F_{\mu \nu}^0\\ F_{\mu \nu}^{\hat{a}}\\ F_{\mu \nu}^N+\frac{1}{2}M^N B_{\mu \nu ,2}\\ \end{array} \right),\\
&-\frac{i}{4}\Theta^{\Lambda m}\tilde{B}_{\mu \nu ,m}\left( F_{\Lambda}^{ \mu \nu}-\frac{1}{4}\Theta _{\Lambda}^{\ n}B^{\mu \nu }_n\right)=-\frac{i}{4}M^N\tilde{B}_{\mu \nu ,2}\left(F^{\mu\nu}_N-\frac{1}{4}E_NB_2^{\mu\nu}\right),
\end{align}
and 
\begin{align}
&D_{\mu}b^2=\partial_{\mu}b^2-(A_{\mu}^0e_2+A_{\mu}^NE_N+A_{\mu N}M^N),\\
&D_{\mu}b^3=\partial_{\mu}b^3-A_{\mu}^0e_3.
\end{align}
Note that there seems $n_v+2$ vectors $A_{\mu}^0, A_{\mu}^{\hat{a}}, A_{\mu}^N,$ and $A_{\mu N}$ at a glance. However, we can gauge away $A_{\mu\nu}^N$ by
\begin{align}
B_{\mu\nu ,2}\rightarrow B_{\mu\nu ,2}-\frac{2}{M^N}F^N_{\mu\nu}, \ \ A_{\mu, N}\rightarrow A_{\mu, N}-\frac{E_N}{M^N}A_{\mu}^N, \label{GCAN}
\end{align}
and eliminate its degree of freedom. Therefore, we obtain $n_v+1$ vectors as usual. Also, $A_{\mu N}$ and $B_{\mu \nu ,2}$ do not have kinetic terms, and should be integrated out.\footnote{ $B_{\mu \nu ,1}$ and $B_{\mu \nu ,3}$ are absent under the parametrization $\eqref{theta_2}$.} As shown in Appendix~\ref{2-form}, eliminating $B_{\mu \nu ,2}$ gives a kinetic term for $A_{\mu N}$. We only show the result here.

After integrating out the two-form field, Eq.~$\eqref{LB}$ at the vacuum is given by \footnote{The vacuum expectation values of the gauge kinetic function are shown in Appendix~\ref{GKF}.} 
\begin{align} 
\nonumber   \mathcal{L}_B=&\frac{1}{4}\langle \hat{\mathcal{I}}_{00}\rangle F^0F^0 +\frac{1}{2}\langle \hat{\mathcal{I}}^N_{\ \ 0}\rangle F_NF^0+\frac{1}{4}\langle\hat{\mathcal{I}}^{NN}\rangle F_NF_N+\frac{1}{4}\langle\hat{\mathcal{I}}_{\hat{a}\hat{b}}\rangle F^{\hat{a}}F^{\hat{b}}\\
&+\left<\frac{1}{2(b^0)^2}\right>\Biggl\{\sum_{u=0}^1\partial_{\mu}b^{u}\partial^{\mu}b^{u}+\left(\partial_{\mu}b^3-e_3A_{\mu}^0\right)^2+\left(\partial_{\mu}b^2-e_2A_{\mu}^0-M^NA_{\mu N}\right)^2\Biggr\},
\end{align}
where
\begin{align}
&\langle \hat{\mathcal{I}}_{00}\rangle=-\frac{\langle e^{- \mathcal{K}}\rangle}{4}\frac{(e_3+{\rm{Re}} \Delta)^2+(e_2+{\rm{Im}} \Delta)^2}{(e_3+{\rm{Re}} \Delta)^2},\\
&\langle \hat{\mathcal{I}}_{\hat{a}\hat{b}}\rangle={\rm{Re}}\left(2c_{\hat{a}\hat{b}}+6\lambda c_{\hat{a}\hat{b}N}\right),\\
&\langle \hat{\mathcal{I}}^N_{\ \ 0}\rangle=-\frac{\langle e^{- \mathcal{K}}\rangle}{4}\frac{M^N(e_2+{\rm{Im}} \Delta)}{(e_3+{\rm{Re}} \Delta)^2},\\
&\langle \hat{\mathcal{I}}^{NN}\rangle=-\frac{\langle e^{- \mathcal{K}}\rangle}{4}\frac{(M^N)^2}{(e_3+{\rm{Re}} \Delta)^2}.
\end{align}
For notational simplicity, here we have omitted the spacetime indices of the field-strength $F_{\mu\nu}$.

It can be found that two hyperscalars, $b^2$ and $b^3$, can be eliminated by
\begin{align}
&A_{\mu}^0\rightarrow A_{\mu}^0 +\frac{1}{e_3}\partial_{\mu}b^3,\label{GCb3}\\
&A'_{\mu}\rightarrow A'_{\mu} +\frac{1}{M^N}\partial_{\mu}b^2. \label{GCb2}
\end{align}
where we have redefined $A_{\mu N}$ as
\begin{align}
A'_{\mu}=A_{\mu N}+\frac{e_2}{M^N}A_{\mu}^0.    
\end{align}
Then, the Lagrangian becomes
\begin{align}
\nonumber \mathcal{L}_B=&\frac{1}{4}\langle \hat{\mathcal{I}}_{00}\rangle F^0F^0 +\frac{1}{2}\langle \hat{\mathcal{I}}^N_{\ \ 0}\rangle F_NF^0+\frac{1}{4}\langle\hat{\mathcal{I}}^{NN}\rangle F_NF_N+\frac{1}{4}\langle\hat{\mathcal{I}}_{\hat{a}\hat{b}}\rangle F^{\hat{a}}F^{\hat{b}}\\
&+\left<\frac{1}{2(b^0)^2}\right>\Biggl\{ \sum_{u=0}^1\partial_{\mu}b^{u}\partial^{\mu}b^{u}+\left(e_3A_{\mu}^0\right)^2+\left(M^NA'_{\mu }\right)^2\Biggr\}.
\end{align}
It contains $n_v-1$ massless vector fields and $2$ massive ones. There are also two massless hyperscalars, $b^0$ and $b^1$. For the massive modes, we can diagonalize their kinetic and mass matrices by 
\begin{align}
\left( \begin{array}{cc} B_{\mu}\\ B'_{\mu}\\ \end{array} \right)  =\frac{\langle e^{- \mathcal{K}/2}\rangle }{2(e_3+{\rm{Re}}\Delta)}\left(\begin{array}{cc}  \sqrt{\mathbb{E}_+}e_3{\rm{cos}}\theta& \sqrt{\mathbb{E}_+}M^N{\rm{sin}}\theta \\  -\sqrt{\mathbb{E}_-}e_3{\rm{sin}}\theta&  \sqrt{\mathbb{E}_-}M^N{\rm{cos}}\theta\\ \end{array} \right) \left( \begin{array}{cc} A^0_{\mu}\\ A'_{\mu}\\ \end{array} \right),    
\end{align}
where 
\begin{align}
&\mathbb{E}_{\pm}= 1+\frac{{\rm{Re}}\Delta}{e_3}+\frac{|\Delta|^2}{2e_3^2}\pm \frac{|\Delta|}{2e_3^2}\sqrt{|\Delta|^2+4e_3{\rm{Re}}\Delta+4e_3^2},\\
&{\rm{tan}}2\theta =\frac{2e_3{\rm{Im}}\Delta}{|\Delta|^2+2e_3{\rm{Re}}\Delta}.
\end{align}
Finally, we obtain
\begin{align}
\nonumber \mathcal{L}_B=&-\frac{1}{4}F(B)F(B) -\frac{1}{4}F(B')F(B')+\frac{1}{4}\left<\mathcal{I}_{\hat{a}\hat{b}} \right>F^{\hat{a}}F^{\hat{b}}\\
&+\left<\frac{1}{2(b^0)^2}\right> \sum_{u=0}^1\partial_{\mu}b^{u}\partial^{\mu}b^{u}+\frac{1}{2}m^2_BB_{\mu}^2+\frac{1}{2}m^2_{B'}B_{\mu}'^2, \label{LB_final}
\end{align}
where $F(B)$ and $F(B')$ denote the field-strength of $B_{\mu}$ and $B'_{\mu}$. Their masses are given by
\begin{align}
&m^2_B=\left<\frac{ e^{\mathcal{K}}}{(b^0)^2}\right>4e_3^2\mathbb{E}_-,\\
&m^2_{B'}=\left<\frac{ e^{\mathcal{K}}}{(b^0)^2}\right>4e_3^2\mathbb{E}_+.
\end{align}
Remarkably, $m_B$ and $m_{B'}$ are related to $m_1$ and $m_2$ by
\begin{align}
&m_B=\left||m_2|-|m_1|\right|,\label{Amb}\\
&m_{B'}=|m_2|+|m_1|.\label{Ambp}
\end{align}
We see some implications of their hierarchical structures.

Finally, let us check the mass of $z^i$. Defining the fluctuation around the vacuum, 
\begin{align}
z^i=\left< z^i\right> + \tilde{z}^i,  
\end{align}
we can expand the scalar potential as 
\begin{align}
V=&  \left<V\right>+\left(\left<\partial_iV\right>\tilde{z}^i+{\rm{h.c.}}\right)+\left<\partial_i\partial_{\bar{j}}V\right>\tilde{z}^i\bar{\tilde{z}}^{\bar{j}}+\left(\frac{1}{2}\left<\partial_i\partial_jV\right>\tilde{z}^i\tilde{z}^j+{\rm{h.c.}}\right)+\cdots,
\end{align}
where the ellipsis denotes higher order couplings of $\tilde{z}^i$. 
The first and the second terms vanish due to the (Minkowski) vacuum conditions.\footnote{In order to avoid the runaway of $b^0$, the vacuum should be the Minkowski one.} The third and fourth terms are expressed as 
\begin{align}
&\left<\partial_{\hat{a}}\partial_{\bar{\hat{b}}}V\right>=\delta_{\hat{a}\hat{b}}\left<\frac{2e^{3\mathcal{K}}}{(b^0)^2}\frac{|C|^2}{G}g^{N\bar{N}}\right>(|\Delta|^2+2e_3{\rm{Re}}\Delta+2e_3^2),\\
&\left<\partial_{\hat{a}}\partial_{\hat{b}}V\right>=-\delta_{\hat{a}\hat{b}}\left<\frac{2e^{3\mathcal{K}}}{(b^0)^2}\frac{C^2}{G} g^{N\bar{N}}\right>\bar{\Delta}(\bar{\Delta}+2e_3),\\
&\left<\partial_{N}\partial_{\bar{N}}V\right>=\left<\partial_{N}\partial_{\bar{\hat{a}}}V\right>=\left<\partial_N\partial_NV\right>=\left<\partial_{N}\partial_{\hat{a}}V\right>=0,
\end{align}
by assuming Eq.~$\eqref{diag}$. Note that $\tilde{z}^N$ is massless. Taking into account the canonical normalization $(\tilde{z}^{\hat{a}}\rightarrow \tilde{z}^{\hat{a}}/G)$, and diagonalizing the mass matrix, we obtain
\begin{align}
V=\frac{1}{2}m^2_{x}(x^{\hat{a}})^2+\frac{1}{2}m^2_{y}(y^{\hat{a}})^2+\cdots ,
\end{align}
where 
\begin{align}
&m_{x}=|m_+|+|m_-|,\label{m_x}\\
&m_{y}=\left||m_+|-|m_-|\right|,\label{m_y}\\
&\left( \begin{array}{cc} x^{\hat{a}}\\ y^{\hat{a}}\\ \end{array} \right)  =\frac{1}{\sqrt{2}}\left(\begin{array}{cc}  {\rm{cos}}\varphi& {\rm{sin}}\varphi \\  -{\rm{sin}}\varphi&  {\rm{cos}}\varphi\\ \end{array} \right) \left( \begin{array}{cc} {\rm{Re}}\tilde{z}^{\hat{a}}\\{\rm{Im}}\tilde{z}^{\hat{a}}\\ \end{array} \right),  \\ 
&{\rm{tan}}2\varphi =-\frac{{\rm{Im}}(m_+m_-)}{2{\rm{Re}}(m_+m_-)}
\end{align}
The mass of $\tilde{z}^{\hat{a}}$ and $\lambda^{\hat{a}\pm }$ are related by Eqs.~$\eqref{m_x}$ and $\eqref{m_y}$ under the choice of Eq.~$\eqref{diag}$. In general cases, this relation does not seem to hold unlike the relations $\eqref{Amb}$ and $\eqref{Ambp}$. For general values of $c_{\hat{a}\hat{b}}$ and $c_{\hat{a}\hat{b}N}$, the gaugino masses (the last line in Eq.~$\eqref{Fmass}$) are given by 
\begin{align}
\left<-\frac{i}{2b^0}e^{3\mathcal{K}/2}\sqrt{g^{N\bar{N}}}f_{\hat{a}\hat{b}N}\right>\bar{\Delta}\bar{\lambda}^{\hat{a}-}\lambda^{\hat{b}-}+\left<-\frac{i}{2b^0}e^{3\mathcal{K}/2}\sqrt{g^{N\bar{N}}}f_{\hat{a}\hat{b}N}\right>(\bar{\Delta}+2e_3)\bar{\lambda}^{\hat{a}+}\lambda^{\hat{b}+},
\end{align}
and the gauginos have different masses depending on $c_{\hat{a}\hat{b}}$ and $c_{\hat{a}\hat{b}N}$. The boson masses of $\tilde{z}^{\hat{a}}$ are also similar.

\subsection{Summary of mass spectrum}\label{summary}
In table~\ref{table_spectrum}, we summarize the spectrum (at the tree level) obtained in the previous subsections, where $\Delta'\equiv \Delta+2e_3$.

\begin{table}[t]
\centering 
  \begin{tabular}{|c|c|} \hline
      field & mass \\ \hline \hline
     $\Psi_+\equiv\{\psi_{\mu +},\eta_{\bullet }\}$ & $|m_1|=\left|\left<  \frac{e^{\mathcal{K}/2}}{b^0}\right>\Delta \right|$  \\ \hline
    $\Psi_-\equiv\{\psi_{\mu -},\tilde{\eta}_{\bullet }\}$ & $|m_2|=\left|\left<  \frac{e^{\mathcal{K}/2}}{b^0}\right>\Delta'\right|$  \\ \hline
    $ B_{\mu}$ & $m_B=\left||m_2|-|m_1|\right|$ \\ \hline
    $ B'_{\mu}$ & $m_{B'}=|m_2|+|m_1|$ \\ \hline
     $\lambda^{\hat{a}-}$ & $|m_-|=\left|\left<  \frac{e^{3\mathcal{K}/2}}{b^0}\sqrt{g^{N\bar{N}}}\frac{C}{G}\right>\bar{\Delta}\right|$  \\ \hline
     $\lambda^{\hat{a}+}$ & $|m_+|=\left|\left<  \frac{e^{3\mathcal{K}/2}}{b^0}\sqrt{g^{N\bar{N}}}\frac{C}{G}\right>\bar{\Delta}'\right|$  \\ \hline
     $x^{\hat{a}}$ & $m_x=|m_+|+|m_-|$  \\ \hline
      $y^{\hat{a}}$ & $m_y=\left||m_+|-|m_-|\right|$  \\ \hline
    $g_{\mu\nu}, z^N, A^{\hat{a}}, b^0, b^1$ & $0$  \\ \hline
  \end{tabular}
  \caption{Mass spectrum} \label{table_spectrum}
\end{table}

As we saw before, the first supersymmetry ($\epsilon_+$) recovers in the limit $\Delta \rightarrow 0$. 
In this case, the first gravitino~$\psi_{\mu +}$ becomes massless ($|m_1|=0$), and it belongs to the $\mathcal{N}=1$ gravitational multiplet with $g_{\mu\nu}$.
The second gravitino $\psi_{\mu -}$ and $\tilde{\eta}_{\bullet }, B_{\mu}, B'_{\mu}$ have degenerate masses ($|m_2|=m_B=m_{B'}$), and they form an $\mathcal{N}=1$ massive spin $3/2$ multiplet. Also, half of the gauginos $\lambda^{\hat{a}-}$ become massless, and with $A^{\hat{a}}$, they form $n_v-1$ massless $\mathcal{N}=1$ vector multiplets. The remaining gauginos $\lambda^{\hat{a}+}$ have the mass $|m_+|$, and are degenerate with $\tilde{z}^{\hat{a}}=\{x^{\hat{a}},y^{\hat{a}}\}$. Then, they form $n_v-1$ massive $\mathcal{N}=1$ chiral multiplets. Finally, $\{\zeta_{-},b^0,b^1\}$ and $\{\lambda^{N+},z^N\}$ form $\mathcal{N}=1$ massless chiral multiplets. 

On the other hand, $\Delta'= 0$ (or $\Delta=-2e_3$) corresponds to the case where the second supersymmetry ($\epsilon_-$) recovers and the first one ($\epsilon_+$) is broken, which can be seen from Eqs.~$\eqref{psipm}$-$\eqref{zetapm}$. 
The mass spectrum is the same as the case of $\Delta=0$, but the roles of $\psi_{\mu -},\tilde{\eta}_\bullet$, and $\lambda^{\hat{a}+}$ are just replaced by $\psi_{\mu +},\eta_\bullet$, and $\lambda^{\hat{a}-}$, respectively. 

Toward a phenomenological application, let us consider a case of hierarchical supersymmetry breaking~$|\Delta'|\gg |\Delta|$. 
Then, the mass spectrum of the massive vectors~$\{B_\mu,B'_\mu\}$ 
and the fermions~$\Psi_-\equiv\{\psi_{\mu -},\tilde{\eta}_\bullet\}$ and $\Psi_+\equiv\{\psi_{\mu +},\eta_\bullet\}$ satisfies 
\begin{align}
m_{B'}\gtrsim|m_2|\gtrsim m_B\gg |m_1|,    \label{ineq}
\end{align}
From the results shown in Appendix~\ref{int}, we find that all the interactions among these fields 
schematically take the following forms: 
\begin{align}
&B\bar{\Psi}_-\Psi_+, \label{BPP} \\
&B'\bar{\Psi}_-\Psi_+.  \label{BpPP}
\end{align}
Through these interactions, the possible decay processes of the heavy fields~$B'_\mu$ and $\Psi_-$ are 
\begin{align}
&\Psi_-\rightarrow \Psi_+, B, \label{P_PB} \\
&B'\rightarrow \Psi_-, \Psi_+.  \label{Bp_PP}
\end{align}
However, due to the mass relations~$\eqref{Amb}$ and $\eqref{Ambp}$, 
their decay rates vanish at least at the tree level. 

As for the other massive fields~$x^{\hat{a}}$, $y^{\hat{a}}$ and $\lambda^{\hat{a}\pm}$, 
there are no specific mass relations since their masses depend on the free parameter~$C$ 
(or $c_{\hat{a}\hat{b}N}$ in general). 
Thus, their decays to the light states in $\Psi_+$ can have nonvanishing rates.

\begin{figure}[t]
\includegraphics[width=15.0cm]{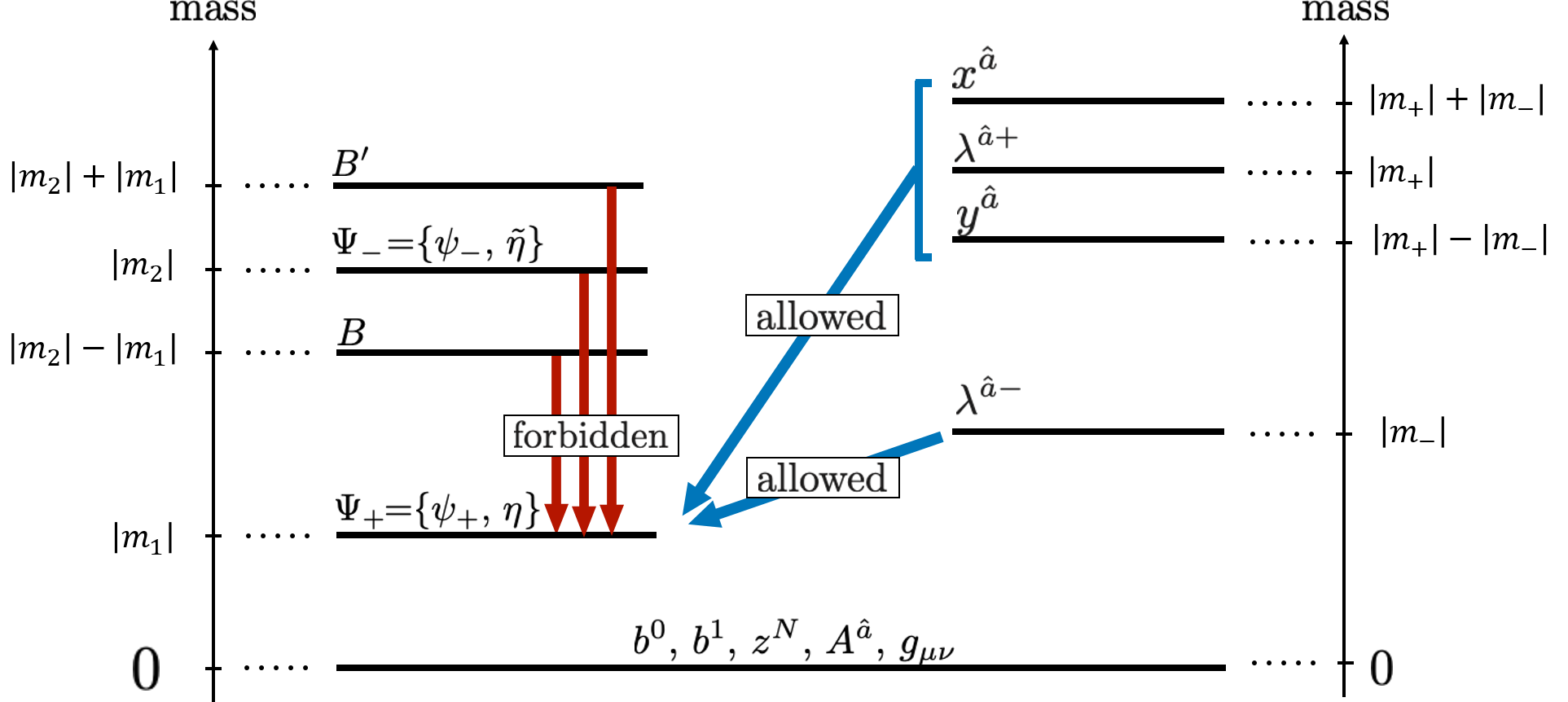}
\caption{Decay modes to $\Psi_+$}
\label{fig_spectrum}
\end{figure}

In summary, we found that the direct decays from the heavy particles~$B, B', \Psi_-$ to the light ones in $\Psi_+$ 
are not allowed (at the tree level), but those from $x^{\hat{a}}, y^{\hat{a}}, \lambda^{\hat{a}\pm}$ to $\Psi_+$ can be allowed depending on $C$. We schematically show these allowed/forbidden decay processes in Fig.~\ref{fig_spectrum}. These restrictions may become important when we discuss the cosmological history based on models of extended supergravity. 

\section{Conclusion}\label{con}
In this paper, we have investigated the $\mathcal{N}=2$ supergravity model 
that interpolates the full and the partial breakings of supersymmetries, which is a generalization of Ref.~\cite{Ferrara:1995gu}. We extend the model by introducing additional vector multiplets. 
As can be inferred from the studies of the partial breaking of extended supergravity~\cite{Ferrara:1995gu,Ferrara:1995xi,Fre:1996js,Louis:2009xd}, the magnetic gauging is important and we chose the embedding tensor as Eq.~$\eqref{theta_2}$ with Eq.~$\eqref{MN}$. Then we found the conditions for an $\mathcal{N}=0$ Minkowski vacuum, 
which is continuously connected to $\mathcal{N}=1$ preserving ones.

The breaking scales of the two supersymmetries can be chosen independently in our model (see Eqs.~$\eqref{psipm}$-$\eqref{zetapm}$).  Thus, we can discuss the case of the cascade supersymmetry breakings in which the two supersymmetry breaking scales are separated hierarchically. This is phenomenologically interesting when we consider the string compactifications. 
In such a case, an approximate $\mathcal{N}=1$ supersymmetry appears at intermediate energies between the two scales. 
In contrast to other phenomenological supersymmetric models, we can quantitatively discuss the effects of the second supersymmetry 
breaking on light modes in the approximate $\mathcal{N}=1$ sector. 

We computed the mass spectrum (see table~\ref{table_spectrum}), which is indeed characterized by two different scales, $|\Delta|$ and $|\Delta'|$. We found that there are non-trivial relations in the spectrum, $\eqref{Amb}$ and~$\eqref{Ambp}$. 
Even for $|\Delta'|\gg |\Delta|$, there are no direct decay processes from the heavy fields ($B,B',\Psi_-$) whose masses are of $\mathcal{O}(|\Delta'|)$ to the first gravitino ($\Psi_+$) in our model (see figure.~\ref{fig_spectrum}). This property may be important when models based on extended supergravity\footnote{It is remarkable that supergravity in higher dimensional spacetime (such as the low energy effective theories of superstrings) inevitably becomes an extended supergravity in four dimensions at low energies, if the supersymmetry-breaking scales are lower than the compactification scales.} are applied to cosmological scenarios.

There are several issues to be addressed: The first one is the generality of the gauging. In this work, we have not discussed the most general gauging, taking the embedding tensor as Eq.~$\eqref{theta_2}$ with the simplification~$\eqref{MN}$. While our model includes that of Ref.~\cite{Ferrara:1995gu} as a special case, it is worth investigating the other gauging and checking how general our result is. Also, we may allow other vacuum solutions without assuming the ansatz~$\eqref{z_assum}$-$\eqref{f'''_assum}$, and include higher-order terms in $z^i$ of the prepotential $f$. It would be also interesting to extend this analysis to the system with multiple hypermultiplets and non-Abelian gauge symmetries as in Ref.~\cite{Itoyama:2006ef}. We leave them for future works.

\subsection*{Acknowledgements}
Authors would like to thank Hajime Otsuka and Yusuke Yamada for useful discussion. H. A. is supported in part by JSPS KAKENHI Grant Number JP16K05330 and also supported by Institute for Advanced Theoretical and Experimental Physics, Waseda University. S.A. is supported in part by a Waseda University Grant for Special Research Projects (Project number: 2018S-141). 

\begin{appendix}
\section{Notations} \label{notation}

\subsection{Index of vector fields}
The index of the vector fields $\Lambda$ is decomposed as in table~\ref{table_indexV}. 
\begin{table}[h]
 
    \begin{center}
  
  \begin{tabular}{|l|} \hline
      $\Lambda=\{0,1\cdots,n_v\}$\\ $\ \ \ =\{0,i\}=\{0,\hat{a},N\}=\{\hat{\Lambda},N\}$ \\ \hline 
     $i=1,2,\cdots,n_v$  \\ \hline
      $\hat{a}=1,2,\cdots,n_v-1$  \\ \hline
       $\hat{\Lambda}=0,1\cdots,n_v-1$  \\ \hline
     \end{tabular}

    \end{center}
    \caption{Index of vector fields}
    \label{table_indexV}
  
\end{table}

\subsection{Spinor notations}
Here, we summarize spinor conventions.

The SU(2) and Sp(2) invariant tensors satisfy
\begin{align}
&\epsilon^{AB}\epsilon_{BC}=-\delta^A_C, \ \ \epsilon^{12}=\epsilon_{12}=1,\\
&\mathbb{C}^{\alpha \beta}\mathbb{C}_{\beta \gamma}=-\delta^{\alpha}_{\gamma},\ \ \mathbb{C}^{12}=\mathbb{C}_{12}=1,
\end{align}
and the indices of SU(2) and Sp(2) vectors are raised and lowered by
\begin{align}
\epsilon_{AB}P^B=P_A,\ \ \epsilon^{AB}P_B=-P^A,\\
\mathbb{C}_{\alpha \beta}P^{\beta}=P_{\alpha},\ \ \mathbb{C}^{\alpha \beta}P_{\beta}=-P^{\alpha}.
\end{align}

The Pauli matrices are $(\sigma^x)_A^{\ B}(x=1,2,3)$ are
\begin{align}
(\sigma^1)_A^{\ B}=\left(\begin{array}{cc}  0&1 \\   1& 0\\ \end{array} \right), \ \ (\sigma^2)_A^{\ B}=\left(\begin{array}{cc}  0&-i \\   i& 0\\ \end{array} \right), \ \ (\sigma^3)_A^{\ B}=\left(\begin{array}{cc}  1&0 \\   0& -1\\ \end{array} \right).   
\end{align}
Their indices are raised and lowered by $\epsilon_{AB}$ and $\epsilon^{AB}$ defined above.

We denote the chirality of the spinors as 
\begin{align}
&\gamma_5\left( \begin{array}{cc} \psi_A\\ \lambda^{iA}\\ \zeta_{\alpha}\\ \chi_{\bullet} (\tilde{\chi}_{\bullet})\\ \eta_{\bullet} (\tilde{\eta}_{\bullet})\\ \end{array} \right)=\left( \begin{array}{cc} \psi_A\\ \lambda^{iA}\\ \zeta_{\alpha}\\ \chi_{\bullet} (\tilde{\chi}_{\bullet})\\ \eta_{\bullet} (\tilde{\eta}_{\bullet})\\ \end{array} \right), \\
&\gamma_5\left( \begin{array}{cc} \psi^A\\ \lambda^{\bar{i}}_{A}\\ \zeta^{\alpha}\\ \chi^{\bullet} (\tilde{\chi}^{\bullet})\\ \eta^{\bullet} (\tilde{\eta}^{\bullet})\\ \end{array} \right)=-\left( \begin{array}{cc} \psi^A\\ \lambda^{\bar{i}}_{A}\\ \zeta^{\alpha}\\ \chi^{\bullet} (\tilde{\chi}^{\bullet})\\ \eta^{\bullet} (\tilde{\eta}^{\bullet})\\ \end{array} \right).
\end{align}


\section{Integrating out two-form field}\label{2-form}
Here we show the process to integrate out the auxiliary two-form field, which is needed for the embedding tensor formalism~\cite{deWit:2002vt,deWit:2005ub}.

The two-form field appears in Eqs.~$\eqref{kin,v},\eqref{L_top}$. Also, it is contained in Eq.~$\eqref{L_Pauli}$ which is necessary
for deriving the interactions in Appendix~\ref{int}. We summarize them as,
\begin{align}
\nonumber  \mathcal{L}_{{\rm{kin}},v}+\mathcal{L}_{\rm{top}}+\mathcal{L}_{\rm{Pauli}}=&\frac{1}{4}  \mathcal{I}_{\Lambda \Sigma}\mathcal{H}_{\mu \nu}^{\Lambda}\mathcal{H}^{\Sigma \mu \nu}+\frac{i}{4}  \mathcal{R}_{\Lambda \Sigma}\mathcal{H}_{\mu \nu}^{\Lambda}\tilde{\mathcal{H}}^{\Sigma \mu \nu}\\
&-\frac{i}{4}\Theta^{\Lambda m}\tilde{B}_{\mu \nu ,m}\left( F_{\Lambda}^{ \mu \nu}-\frac{1}{4}\Theta _{\Lambda}^{\ n}B^{\mu \nu }_n\right) +\mathcal{H}_{\mu \nu}^{\Lambda}\mathcal{Q}_{\Lambda}^{\mu\nu}, \label{includingB}
\end{align}
where $\mathcal{Q}_{\Lambda}^{\mu\nu}$ comes from Eq.~$\eqref{L_Pauli}$, and is defined by  
\begin{align}
\mathcal{Q}_{\Lambda}^{\mu\nu} \equiv &{\rm{Re}} \mathcal{S}_{\Lambda}^{\mu\nu}+ * i{\rm{Im}} \mathcal{S}_{\Lambda}^{\mu\nu},\\
\nonumber \mathcal{S}_{\Lambda}^{\mu\nu} \equiv & \mathcal{I}_{\Lambda\Sigma}\biggl\{ 2L^{\Sigma}\bar{\psi}^{A\mu}\psi^{B\nu}\epsilon_{AB}-2i\bar{f}^{\Sigma}_{\bar{i}}\bar{\lambda}^{\bar{i}}_A\gamma^{\nu}\psi_B^{\mu}\epsilon^{AB}\\
&+\frac{1}{4}\mathcal{D}_if_j^{\Sigma}\bar{\lambda}^{iA}\gamma^{\mu\nu}\lambda^{jB}\epsilon^{AB}-\frac{1}{2}L^{\Sigma}\bar{\zeta}_{\alpha}\gamma^{\mu\nu}\zeta_{\beta}\mathbb{C}^{\alpha\beta}\biggr\}.
\end{align}
Here $*$ denotes a Hodge dual defined by $*T_{\mu\nu}=-\frac{i}{2}\varepsilon_{\mu\nu\rho\sigma}T^{\rho\sigma}$.

In the gauge $\eqref{GCAN}$, the E.O.M of $B_{\mu\nu, 2}$ yields 
\begin{align}
 \mathcal{I}_{\hat{\Lambda} N}\mathcal{H}_{\mu \nu}^{\hat{\Lambda}}+i  \mathcal{R}_{\hat{\Lambda} N}\tilde{\mathcal{H}}^{\hat{\Lambda}}_ {\mu \nu}-i\tilde{F}_{\mu \nu N}+\frac{i}{2}E_N\tilde{B}_{\mu\nu ,2}+2\mathcal{Q}_{N\mu\nu}=0,    
\end{align}
where we have introduced an index $\hat{\Lambda}=\{0,\hat{a}\}$.
This can be solved as
\begin{align}
&B_{\mu\nu, 2}=-\frac{2}{M^N}\frac{  \mathcal{I}_{N N}}{\mathcal{I}^2_{NN}+\mathcal{P}_{NN}^2}\mathcal{J}_{\mu\nu}+\frac{2i}{M^N}\frac{\mathcal{P}_N}{\mathcal{I}^2_{NN}+\mathcal{P}_{NN}^2}\tilde{\mathcal{J}}_{\mu\nu},\\
&\mathcal{P}_{NN}\equiv \frac{E_N}{M^N}+\mathcal{R}_{NN},\\
&\mathcal{J}_{\mu\nu}\equiv  \mathcal{I}_{N \hat{\Lambda}}F^{\hat{\Lambda}}_{\mu\nu}+i  \mathcal{R}_{N \hat{\Lambda}}\tilde{F}^{\hat{\Lambda}}_{\mu\nu}-i\tilde{F}_{\mu\nu N}+2\mathcal{Q}_N.
\end{align}
By substituting the solution into the Lagrangian~$\eqref{includingB}$, we obtain
\begin{align}
\nonumber \mathcal{L}_{v,kin}+\mathcal{L}_{\rm{top}}+\mathcal{L}_{\rm{Pauli}}=&\frac{1}{4}\hat{\mathcal{I}}_{\hat{\Lambda} \hat{\Sigma}}F^{\hat{\Lambda}}F^{\hat{\Sigma}} +\frac{1}{2}\hat{\mathcal{I}}^N_{\ \ \hat{\Sigma}}F_NF^{\hat{\Sigma}}+\frac{1}{4}\hat{\mathcal{I}}^{NN}F_NF_N\\
\nonumber &+\frac{i}{4}\hat{\mathcal{R}}_{\hat{\Lambda} \hat{\Sigma}}F^{\hat{\Lambda}}\tilde{F}^{\hat{\Sigma}} +\frac{i}{2}\hat{\mathcal{R}}^N_{\ \ \hat{\Sigma}}F_N\tilde{F}^{\hat{\Sigma}}+\frac{i}{4}\hat{\mathcal{R}}^{NN}F_N\tilde{F}_N\\
&+\Sigma^{N}\mathcal{Q}_N+\mathcal{H}^{\hat{\Lambda}}\mathcal{Q}_{\hat{\Lambda}}+{\rm{four\ fermi\ couplings}}, \label{vint}
\end{align}
where
\begin{align}
\hat{\mathcal{I}}_{\hat{\Lambda} \hat{\Sigma}}=&\mathcal{I}_{\hat{\Lambda} \hat{\Sigma}}+\frac{\mathcal{I}_{NN}}{\mathcal{I}^2_{NN}+\mathcal{P}_{NN}^2}\biggl[\mathcal{R}_{\hat{\Lambda} N}\mathcal{R}_{\hat{\Sigma} N}-\mathcal{I}_{\hat{\Lambda} N}\mathcal{I}_{\hat{\Sigma} N}-2\mathcal{I}_{NN}^{-1}\mathcal{P}_{NN}\mathcal{R}_{(\hat{\Lambda} N}\mathcal{I}_{\hat{\Sigma} )N}\biggr],\label{hatILS}\\
\hat{\mathcal{I}}^{N}_{\ \ \hat{\Lambda}}=&\frac{\mathcal{I}_{NN}}{\mathcal{I}^2_{NN}+\mathcal{P}_{NN}^2}\biggl[ -\mathcal{R}_{N\hat{\Lambda} }+\mathcal{P}_{NN}\mathcal{I}_{NN}^{-1}\mathcal{I}_{N\hat{\Lambda}}\biggr],\\
\hat{\mathcal{I}}^{NN}=&\frac{\mathcal{I}_{NN}}{\mathcal{I}^2_{NN}+\mathcal{P}_{NN}^2},\\
\hat{\mathcal{R}}_{\hat{\Lambda} \hat{\Sigma}}=&\mathcal{R}_{\hat{\Lambda} \hat{\Sigma}}+\frac{\mathcal{I}_{NN}}{\mathcal{I}^2_{NN}+\mathcal{P}_{NN}^2}\biggl[\mathcal{I}_{NN}^{-1}\mathcal{P}_{NN}\left(\mathcal{I}_{\hat{\Lambda} N}\mathcal{I}_{\Sigma N}-\mathcal{R}_{\hat{\Lambda} N}\mathcal{R}_{\hat{\Sigma} N}\right)-2\mathcal{R}_{(\hat{\Lambda} N}\mathcal{I}_{\hat{\Sigma} )N}\biggr],\\
\hat{\mathcal{R}}^{N}_{\ \ \hat{\Lambda}}=&\frac{\mathcal{I}_{NN}}{\mathcal{I}^2_{NN}+\mathcal{P}_{NN}^2}\biggl[ \mathcal{I}_{N\hat{\Lambda} }+\mathcal{P}_{NN}\mathcal{I}_{NN}^{-1}\mathcal{R}_{N\hat{\Lambda}}\biggr],\\
\hat{\mathcal{R}}^{NN}=&-\frac{\mathcal{P}_{NN}}{\mathcal{I}^2_{NN}+\mathcal{P}_{NN}^2},\\
\nonumber \Sigma^N=&-\frac{\mathcal{I}_{NN}}{\mathcal{I}^2_{NN}+\mathcal{P}_{NN}^2}\biggl[-i\tilde{F}_N+\mathcal{I}_{N\hat{\Lambda}}F^{\hat{\Lambda}}+i\mathcal{R}_{N\hat{\Lambda}}\tilde{F}^{\hat{\Lambda}}\\
&-i\mathcal{I}_{NN}^{-1}\mathcal{P}_{NN}\left(-iF_N+\mathcal{I}_{N\hat{\Lambda}}\tilde{F}^{\hat{\Lambda}}+i\mathcal{R}_{N\hat{\Lambda}}F^{\hat{\Lambda}}\right)\biggr].\label{def_Sigma}
\end{align}
The first and the second line in Eq.~$\eqref{vint}$ describe the kinetic terms and {\it theta} couplings of physical vector fields. The third line shows the interactions including the vector and the fermion biliner couplings (we have neglected four fermi interactions), which are used in Appendix~\ref{int}.

\section{Expectation values of gauge kinetic function}\label{GKF}
Here we list the explicit expressions of the vacuum expectation values of the gauge kinetic function and the couplings appearing in Eq.~$\eqref{vint}$.

At the vacuum, the gauge kinetic function $\eqref{def_GK}$ is evaluated as, 
\begin{align}
&\langle \mathcal{I}_{00}\rangle = -\frac{\langle \mathcal{K}_0\rangle}{2}\frac{1+(\lambda+\bar{\lambda})\langle \partial_N\mathcal{K}\rangle+2|\lambda|^2\langle \partial_N\mathcal{K}\rangle^2}{1-(\lambda-\bar{\lambda})^2\langle \partial_N\mathcal{K}\rangle^2},\\
&\langle \mathcal{I}_{0N}\rangle = \frac{\langle \mathcal{K}_0\partial_N\mathcal{K}\rangle}{2}\frac{1+(\lambda+\bar{\lambda})\langle \partial_N\mathcal{K}\rangle}{1-(\lambda-\bar{\lambda})^2\langle \partial_N\mathcal{K}\rangle^2},\\
&\langle \mathcal{I}_{NN}\rangle =-\langle \mathcal{K}_0\rangle\frac{\langle \partial_N\mathcal{K}\rangle^2}{1-(\lambda-\bar{\lambda})^2\langle \partial_N\mathcal{K}\rangle^2},\\
&\langle \mathcal{I}_{\hat{a}\hat{b}}\rangle ={\rm{Re}}\left(2c_{\hat{a}\hat{b}}+6\lambda c_{\hat{a}\hat{b}N}\right), \ \ \langle \mathcal{I}_{0\hat{a}}\rangle =\langle \mathcal{I}_{N\hat{a}}\rangle =0,\\
&\langle \mathcal{R}_{00}\rangle = 2{\rm{Im}}c_0-\frac{i\langle \mathcal{K}_0\rangle (\lambda-\bar{\lambda})\langle \partial_N\mathcal{K}\rangle^2}{2}\frac{\lambda+\bar{\lambda}+(\lambda^2+\bar{\lambda}^2)\langle \partial_N\mathcal{K}\rangle}{1-(\lambda-\bar{\lambda})^2\langle \partial_N\mathcal{K}\rangle^2},\\
&\langle \mathcal{R}_{0N}\rangle =-\frac{e_2+{\rm{Im}}\Delta}{M^N}+ \frac{i\langle \mathcal{K}_0\rangle (\lambda-\bar{\lambda})\langle \partial_N\mathcal{K}\rangle^2}{2}\frac{1+(\lambda+\bar{\lambda})\langle \partial_N\mathcal{K}\rangle}{1-(\lambda-\bar{\lambda})^2\langle \partial_N\mathcal{K}\rangle^2},\\
&\langle \mathcal{R}_{NN}\rangle =-\frac{E_N}{M^N}-i\langle \mathcal{K}_0\rangle\frac{(\lambda-\bar{\lambda})\langle \partial_N\mathcal{K}\rangle^3}{1-(\lambda-\bar{\lambda})^2\langle \partial_N\mathcal{K}\rangle^2},\\
&\langle \mathcal{R}_{\hat{a}\hat{b}}\rangle ={\rm{Im}} \left(2c_{\hat{a}\hat{b}}+6\lambda c_{\hat{a}\hat{b}N}\right), \ \ \langle \mathcal{R}_{0\hat{a}}\rangle =\langle \mathcal{R}_{N\hat{a}}\rangle =0.
\end{align}
By substituting these equations into Eqs.~$\eqref{hatILS}$-$\eqref{def_Sigma}$, we obtain
\begin{align}
\langle \hat{\mathcal{I}}_{00}\rangle=&-\frac{\langle e^{- \mathcal{K}}\rangle}{4}\frac{(e_3+{\rm{Re}} \Delta)^2+(e_2+{\rm{Im}} \Delta)^2}{(e_3+{\rm{Re}} \Delta)^2},\\
\langle \hat{\mathcal{I}}_{\hat{a}\hat{b}}\rangle=&\langle \mathcal{I}_{\hat{a}\hat{b}}\rangle,\\
\langle \hat{\mathcal{I}}^N_{\ \ 0}\rangle=&-\frac{\langle e^{- \mathcal{K}}\rangle}{4}\frac{M^N(e_2+{\rm{Im}} \Delta)}{(e_3+{\rm{Re}} \Delta)^2},\\
\langle \hat{\mathcal{I}}^{NN}\rangle=&-\frac{\langle e^{- \mathcal{K}}\rangle}{4}\frac{(M^N)^2}{(e_3+{\rm{Re}} \Delta)^2},\\
\langle \hat{\mathcal{I}}_{0\hat{a}}\rangle=&\langle \hat{\mathcal{I}}^{N}_{\ \ \hat{a}}\rangle=0, \\
\nonumber \langle \Sigma^N \rangle=&\left<e^{-\mathcal{K}}\right>\frac{(M^N)^2}{4(e_3+{\rm{Re}}\Delta)^2}\biggl[ -i\tilde{F}_N+2\left<\partial_N\mathcal{K}\right>{\rm{Im}}\lambda F_N-\frac{i(e_2+{\rm{Im}}\Delta)}{M^N}\tilde{F}^0\\
&+\frac{1}{M^N}\biggl\{e_3+{\rm{Re}}\Delta+2\left<\partial_N\mathcal{K}\right>\left((e_3+{\rm{Re}}\Delta){\rm{Re}}\lambda+(e_2+{\rm{Im}}\Delta){\rm{Im}}\Delta\right)\biggr\}F^0\biggr]
\end{align}
Here we have omitted the expressions of $\langle \hat{\mathcal{R}}\rangle$, which are not necessary in the main text.

\section{Interaction}\label{int}
Here, we calculate the interaction terms up to three point couplings. We focus on the interactions of the massive fermions including two gravitinos.

Before specifying the interactions on the vacuum, here we enumerate the undefined quantities in subsection~\ref{AST} : The covariant derivatives of the fermions, $\mathcal{L}_{\rm{Pauli}}$, and $\mathcal{L}_{\rm{der}}$.

The covariant derivatives of the fermions are given by~\cite{Andrianopoli:1996cm,DallAgata:2003sjo,DAuria:2004yjt,Andrianopoli:2011zj},
\begin{align}
&\mathcal{D}_{\mu}\psi_{A\nu}=\nabla_{\mu} \psi_{A\nu}+\frac{i}{2}\hat{Q}_{\mu}\psi_{A\nu}+\hat{\omega}_{A\mu}^{\ B}\psi_{B\nu},\label{Dpsi}\\
&\mathcal{D}_{\mu}\lambda^{iA}=\nabla_{\mu}\lambda^{iA}-\frac{i}{2}\hat{Q}_{\mu}\lambda^{iA}+\hat{\Gamma}^i_{\ j \mu}\lambda^{jA}+\hat{\omega}_{\ B\mu}^{A}\lambda^{iB},\\
&\mathcal{D}_{\mu}\zeta_{\alpha}=\nabla_{\mu}\zeta_{\alpha}-\frac{i}{2}\hat{Q}_{\mu}\zeta_{\alpha}+\hat{\Delta}_{\alpha \mu}^{\ \beta}\zeta_{\beta},\label{Dzeta}
\end{align}
where $\nabla_{\mu}$ denotes the one including the Levi-Civita connection on spacetime, and $\hat{\Gamma}^i_{\ j \mu}=\Gamma^i_{jk}\partial_{\mu}z^k$. $\hat{Q}_{\mu},\hat{\omega}_{A\mu}^{\ B},$ and $\hat{\Delta}_{ \mu}^{\alpha  \beta}$ are the U(1), SU(2), and Sp(2) (gauged) connections, which are given by
\begin{align}
&\hat{Q}_{\mu}=-\frac{i}{2}\left(\partial_i\mathcal{K}\partial_{\mu}z^i-\partial_{\bar{i}}\mathcal{K}\partial_{\mu}\bar{z}^{\bar{i}}\right),\\
&\hat{\omega}_{A\mu}^{\ B}=\hat{\Delta}_{\alpha \mu}^{\  \beta}=\frac{i}{2b^0}\left(\begin{array}{cc}  \partial_{\mu}b^3-A_{\mu}^0e_3& \partial_{\mu}b^1-i\partial_{\mu}b^2+iA'_{\mu }M^N \\  \partial_{\mu}b^1+i\partial_{\mu}b^2-iA'_{\mu }M^N& -\partial_{\mu}b^3+A_{\mu}^0e_3\\ \end{array} \right) .
\end{align}

$\mathcal{L}_{\rm{Pauli}}$ is given by
\begin{align}
\nonumber \mathcal{L}_{\rm{Pauli}}=&\mathcal{H}_{\mu\nu}^{+\Lambda}\mathcal{I}_{\Lambda\Sigma}\biggl\{ 2L^{\Sigma}\bar{\psi}^{A\mu}\psi^{B\nu}\epsilon_{AB}-2i\bar{f}^{\Sigma}_{\bar{i}}\bar{\lambda}^{\bar{i}}_A\gamma^{\nu}\psi_B^{\mu}\epsilon^{AB}\\
&+\frac{1}{4}\mathcal{D}_if_j^{\Sigma}\bar{\lambda}^{\bar{i}A}\gamma^{\mu\nu}\lambda^{jB}\epsilon^{AB}-\frac{1}{2}L^{\Sigma}\bar{\zeta}_{\alpha}\gamma^{\mu\nu}\zeta_{\beta}\mathbb{C}^{\alpha\beta}\biggr\}+{\rm{h.c.}}, \label{L_Pauli}
\end{align}
where $\mathcal{H}^{\pm}_{\mu\nu}\equiv \frac{1}{2}(\mathcal{H}_{\mu\nu}\pm \tilde{\mathcal{H}}_{\mu\nu})$ are (anti) self-dual tensors. 

Finally, $\mathcal{L}_{\rm{der}}$ is given by
\begin{align}
\nonumber \mathcal{L}_{\rm{der}}=&-g_{i\bar{j}}\partial_{\mu}\bar{z}^{\bar{j}}\bar{\psi}_A^{\mu}\lambda^{iA}-2\mathcal{U}_u^{A\alpha}D_{\mu}b^u\bar{\psi}_A^{\mu}\zeta_{\alpha}+g_{i\bar{j}}\partial_{\mu}\bar{z}^{\bar{j}}\bar{\lambda}^{iA}\gamma^{\mu\nu}\psi_{A\nu}\\
&+2\mathcal{U}_u^{A\alpha}D_{\mu}b^u\bar{\zeta}_{\alpha}\gamma^{\mu\nu}\psi_{A\nu}+{\rm{h.c.}}. \label{L_der}
\end{align}

Then, let us see the explicit forms of the interactions on the vacuum and under the gauge conditions~$\chi=\tilde{\chi}=0$, Eqs.~$\eqref{GCAN}, \eqref{GCb3}$, and $\eqref{GCb3}$.

From Eq.~$\eqref{L_Pauli}$, we can obtain the vector and fermion  bilinear couplings. It also contains the two-form field $B_{\mu\nu,2}$ and affects the process of integrating out $B_{\mu\nu,2}$ (see Appendix~\ref{2-form}). After integrating out $B_{\mu\nu,2}$, $\mathcal{L}_{B}+\mathcal{L}_{\rm{Pauli}}$ produces    
\begin{align}
\nonumber  \mathcal{L}_{{\rm{int}},1}=&\mathcal{O}_1F_{\mu\nu}^+(B)\bar{\psi}^{+\mu}\psi^{-\nu}+\mathcal{O}_2F_{\mu\nu}^+(B)\left(\bar{\eta}^{\bullet}\gamma^{\nu}\psi^{\mu}_--\bar{\tilde{\eta}}^{\bullet}\gamma^{\nu}\psi^{\mu}_+\right)+\mathcal{O}_3F_{\mu\nu}^+(B)\bar{\tilde{\eta}}_{\bullet}\gamma^{\mu\nu}\eta_{\bullet}\\
\nonumber  &+\mathcal{O}_{4\hat{a}\hat{b}}F_{\mu\nu}^+(B)\left(-\bar{\lambda}^{\hat{a}+}\gamma^{\mu\nu}\lambda^{\hat{b}-}+\bar{\lambda}^{\hat{a}-}\gamma^{\mu\nu}\lambda^{\hat{b}+}\right)\\
\nonumber  &+\mathcal{O}_5F_{\mu\nu}^+(B')\bar{\psi}^{+\mu}\psi^{-\nu}+\mathcal{O}_6F_{\mu\nu}^+(B')\left(\bar{\eta}^{\bullet}\gamma^{\nu}\psi^{\mu}_--\bar{\tilde{\eta}}^{\bullet}\gamma^{\nu}\psi^{\mu}_+\right)+\mathcal{O}_7F_{\mu\nu}^+(B')\bar{\tilde{\eta}}_{\bullet}\gamma^{\mu\nu}\eta_{\bullet}\\
\nonumber  &+\mathcal{O}_{8\hat{a}\hat{b}}F_{\mu\nu}^+(B')\left(-\bar{\lambda}^{\hat{a}+}\gamma^{\mu\nu}\lambda^{\hat{b}-}+\bar{\lambda}^{\hat{a}-}\gamma^{\mu\nu}\lambda^{\hat{b}+}\right)\\
\nonumber &-2i\langle e^{\mathcal{K}/2}\rangle \left<\mathcal{I}_{\hat{a}\hat{b}}\right> F^{\hat{a}+}_{\mu\nu}\left(-\bar{\lambda}^{\bar{\hat{b}}}_+\gamma^{\nu}\psi^{\mu}_-+\bar{\lambda}^{\bar{\hat{b}}}_-\gamma^{\nu}\psi^{\mu}_+\right)\\
&-  \frac{2\left<e^{\mathcal{K}}\right>}{3}{\rm{Re}}\lambda \left<\mathcal{I}_{\hat{a}\hat{b}}\right>\left<  f_{\hat{c}\hat{d}N}g^{\hat{c}\bar{\hat{b}}}\right>F^{\hat{a}+}_{\mu\nu}\left(-\bar{\lambda}^{\hat{d}+}\gamma^{\mu\nu}\tilde{\eta}_{\bullet}+\bar{\lambda}^{\hat{d}-}\gamma^{\mu\nu}\eta_{\bullet}\right)+{\rm{h.c.}},\label{Lint1}
\end{align}
in addition to Eq.~$\eqref{LB_final}$. Here we have omitted four fermi interactions which come from the integration of $B_{\mu\nu ,2}$. The coefficients are given by 
\begin{align}
&\mathcal{O}_1=\left<  \frac{4e^{\mathcal{K}/2}}{b^0}\right>\frac{1}{|m_2|+|m_1|}\left(e_3e^{i\theta} +\Delta {\rm{cos}}\theta \right),\\
&\mathcal{O}_2=\left<  -\frac{4ie^{\mathcal{K}/2}}{3b^0}\right>\frac{1}{|m_2|+|m_1|}\left(e_3e^{-i\theta} +\bar{\Delta} {\rm{cos}}\theta \right),\\
&\mathcal{O}_3=\left<  \frac{e^{\mathcal{K}/2}}{9b^0}\right>\frac{1}{|m_2|+|m_1|}\left(e_3e^{i\theta} +\Delta {\rm{cos}}\theta\right),\\
&\mathcal{O}_{4\hat{a}\hat{b}}=\left<  \frac{e^{\mathcal{K}}}{8b^0}\right>\left<  f_{\hat{a}\hat{b}N}g^{N\bar{N}}\right>\frac{1}{{\rm{Re}}\lambda(|m_2|+|m_1|)}\left(e_3e^{-i\theta} +\bar{\Delta} {\rm{cos}}\theta\right),
\end{align}
$\mathcal{O}_{5,6,7,8}$ are obtained by the replacements $\mathcal{O}_{1,2,3,4}$ with ${\rm{cos}}\theta\rightarrow -{\rm{sin}}\theta, {\rm{sin}}\theta\rightarrow {\rm{cos}}\theta$, and $|m_1|\rightarrow -|m_1|$ in their expressions. In the derivation, we have assumed ${\rm{Re}c_0}=0$.

From Eq.~$\eqref{L_der}$, we obtain 
\begin{align}
\nonumber \mathcal{L}_{{\rm{int}},2}=&\frac{1}{3{\rm{Re}}\lambda}\partial_{\mu}\bar{\tilde{z}}^{\bar{N}}\left(\bar{\eta}_{\bullet}\gamma^{\nu}\gamma^{\mu}\psi_{+\nu}+\bar{\tilde{\eta}}_{\bullet}\gamma^{\nu}\gamma^{\mu}\psi_{-\nu}\right) -\langle g_{\hat{a}\bar{\hat{b}}}\rangle \partial_{\mu}\bar{\tilde{z}}^{\bar{\hat{b}}}\left(\bar{\lambda}^{\hat{a}+}\gamma^{\nu}\gamma^{\mu}\psi_{+\nu}+\bar{\lambda}^{\hat{a}-}\gamma^{\nu}\gamma^{\mu}\psi_{-\nu}\right)\\
\nonumber &+\left<\frac{1}{3b^0}\right>\partial_{\mu}b^0\left(\bar{\eta}_{\bullet}\gamma^{\nu}\gamma^{\mu}\psi_{+\nu}+\bar{\tilde{\eta}}_{\bullet}\gamma^{\nu}\gamma^{\mu}\psi_{-\nu}\right)+\left<\frac{i}{3b^0}\right>\partial_{\mu}b^1\left(\bar{\eta}_{\bullet}\gamma^{\nu}\gamma^{\mu}\psi_{+\nu}-\bar{\tilde{\eta}}_{\bullet}\gamma^{\nu}\gamma^{\mu}\psi_{-\nu}\right)\\
\nonumber &+\left<\frac{1}{3(b^0)^2}\right>\frac{M^Ne_3}{{\rm{Re}\lambda} (|m_2|+|m_1|)}B_{\mu}\left(e^{-i\theta}\bar{\tilde{\eta}}_{\bullet}\gamma^{\nu}\gamma^{\mu}\psi_{+\nu}+e^{i\theta}\bar{\eta}_{\bullet}\gamma^{\nu}\gamma^{\mu}\psi_{-\nu}\right)\\
&+\left<\frac{1}{3(b^0)^2}\right>\frac{M^Ne_3}{{\rm{Re}\lambda }(|m_2|-|m_1|)}B'_{\mu}\left(e^{-i\theta}\bar{\tilde{\eta}}_{\bullet}\gamma^{\nu}\gamma^{\mu}\psi_{+\nu}+e^{i\theta}\bar{\eta}_{\bullet}\gamma^{\nu}\gamma^{\mu}\psi_{-\nu}\right)+{\rm{h.c.}}.\label{Lint2}
\end{align}

Also, we have the scalar and the fermion bilinear couplings from Eq.~$\eqref{F_mass}$, 
\begin{align}
\nonumber \mathcal{L}_{{\rm{int}},3}=&2\left(\tilde{\mathbb{S}}_{11}+\tilde{\mathbb{S}}_{12}\right)\bar{\psi}^+_{\mu}\gamma^{\mu\nu}\psi_{\nu}^++2\left(\tilde{\mathbb{S}}_{11}-\tilde{\mathbb{S}}_{12}\right)\bar{\psi}^-_{\mu}\gamma^{\mu\nu}\psi_{\nu}^-\\
\nonumber &+\biggl\{i\left<g_{\hat{a}\bar{\hat{b}}}\right>\left(\tilde{W}^{\hat{a}11}+\tilde{W}^{\hat{a}12}\right)\bar{\lambda}_+^{\bar{\hat{b}}}\gamma_{\mu}\psi_+^{\mu}+\frac{2i\left<\partial_N \mathcal{K}\right>}{3}\left(\tilde{W}^{N11}+\tilde{W}^{N12}\right)\bar{\eta}^{\bullet}\gamma_{\mu}\psi_+^{\mu}\\
\nonumber &+i\tilde{g}_{N\bar{\hat{a}}}\left<g^{N\bar{N}}\partial_N \mathcal{K}\right>\bar{m}_1\bar{\lambda}_+^{\bar{\hat{a}}}\gamma_{\mu}\psi_+^{\mu}+\frac{2i}{3}\left<g^{N\bar{N}}\right>\tilde{g}_{N\bar{N}}\bar{m}_1\bar{\eta}^{\bullet}\gamma_{\mu}\psi_+^{\mu}\\
\nonumber &+(+\rightarrow -, \tilde{W}^{i12}\rightarrow -\tilde{W}^{i12},m_1\rightarrow m_2)\biggr\}-\frac{2i}{3}\left(\tilde{N}_1^{\ 1}+\tilde{N}_2^{\ 1}\right)\bar{\tilde{\eta}}^{\bullet}\gamma_{\mu}\psi_-^{\mu}\\
\nonumber &+\frac{2i}{3}\left(\tilde{N}_1^{\ 1}-\tilde{N}_2^{\ 1}\right)\bar{\eta}^{\bullet}\gamma_{\mu}\psi_+^{\mu}+\frac{1}{9}\left(\tilde{\mathcal{M}}^{1 1}+\tilde{\mathcal{M}}^{1 2}\right)\bar{\tilde{\eta}}_{\bullet}\tilde{\eta}_{\bullet}+\frac{1}{9}\left(\tilde{\mathcal{M}}^{1 1}-\tilde{\mathcal{M}}^{1 2}\right)\bar{\eta}_{\bullet}\eta_{\bullet}\\
\nonumber &+\biggl\{ -\frac{1}{3}\left(\tilde{\mathcal{M}}^1_{\hat{a}1}+\tilde{\mathcal{M}}^2_{\hat{a}1}\right)\bar{\tilde{\eta}}_{\bullet}\lambda^{\hat{a}-}-\frac{2}{9\left<\partial_N \mathcal{K}\right>}\left(\tilde{\mathcal{M}}^1_{N1}+\tilde{\mathcal{M}}^2_{N1}\right)\bar{\tilde{\eta}}_{\bullet}\tilde{\eta}_{\bullet}\\
\nonumber &-(-\rightarrow +, \tilde{\mathcal{M}}^2_{i1}\rightarrow -\tilde{\mathcal{M}}^2_{i1})\biggr\}+\biggl\{\left(\tilde{M}_{11\hat{a}\hat{b}}+\tilde{M}_{12\hat{a}\hat{b}}\right)\bar{\lambda}^{\hat{a}+}\lambda^{\hat{b}+}\\
\nonumber &+\frac{4}{3\left<\partial_N \mathcal{K}\right>}\left(\tilde{M}_{11\hat{a}N}+\tilde{M}_{12\hat{a}N}\right)\bar{\lambda}^{\hat{a}+}\eta_{\bullet}+\frac{4}{9}\left<g^{N\bar{N}}\right>\left(\tilde{M}_{11NN}+\tilde{M}_{12NN}\right)\bar{\eta}_{\bullet}\eta_{\bullet}\\
&+(+\rightarrow -, \tilde{M}_{12ij}\rightarrow -\tilde{M}_{12ij})\biggr\}+{\rm{h.c.}},\label{Lint3}
\end{align}
where we have defined the fluctuations from the expectation values of Eqs.~$\eqref{psipsi}$-$\eqref{zz}$ and $g_{i\bar{j}}$, and distinguished them by adding tilde. They are the functions of $\tilde{z}^i$ and $b^0$. $+(-)\rightarrow -(+)$ denotes that $\psi_+(\psi_-)$ and $\eta(\tilde{\eta})$ should be replaced by $\psi_-(\psi_+)$ and $\tilde{\eta}(\eta)$ respectively.

Finally, the covariant derivatives in Eqs.~$\eqref{Dpsi}$-$\eqref{Dzeta}$ produce the following terms,
\begin{align}
\nonumber \mathcal{L}_{{\rm{int}},4}=&-\frac{i}{2}\frac{\varepsilon^{\mu\nu\rho\sigma}}{\sqrt{-g}}\tilde{\hat{Q}}_{\rho}\left(\bar{\psi}^+_{\mu}\gamma_{\nu}\psi_{+\sigma}+\bar{\psi}^-_{\mu}\gamma_{\nu}\psi_{-\sigma}\right)+\left<\frac{i}{2b^0}\right>\frac{\varepsilon^{\mu\nu\rho\sigma}}{\sqrt{-g}}\partial_{\rho}b^1\left(\bar{\psi}^+_{\mu}\gamma_{\nu}\psi_{+\sigma}-\bar{\psi}^-_{\mu}\gamma_{\nu}\psi_{-\sigma}\right)\\
\nonumber &+\left<\frac{i}{2(b^0)^2}\right>\frac{\varepsilon^{\mu\nu\rho\sigma}}{\sqrt{-g}} \frac{e_3m^N}{{\rm{Re}}\lambda (|m_2|+|m_1|)}B_{\rho}\left(e^{i\theta}\bar{\psi}^+_{\mu}\gamma_{\nu}\psi_{-\sigma}+e^{-i\theta}\bar{\psi}^-_{\mu}\gamma_{\nu}\psi_{+\sigma}\right)\\
\nonumber &-\left<\frac{1}{2(b^0)^2}\right> \frac{\varepsilon^{\mu\nu\rho\sigma}}{\sqrt{-g}}\frac{e_3m^N}{{\rm{Re}}\lambda (|m_2|-|m_1|)}B'_{\rho}\left(e^{i\theta}\bar{\psi}^+_{\mu}\gamma_{\nu}\psi_{-\sigma}-e^{-i\theta}\bar{\psi}^-_{\mu}\gamma_{\nu}\psi_{+\sigma}\right)\\
\nonumber &+\biggl\{-\frac{1}{4} \left<g_{\hat{a}\bar{\hat{b}}}\right>\bar{\lambda}_+^{\bar{\hat{b}}}\gamma^{\mu}\tilde{\hat{Q}}_{\mu}\lambda^{\hat{a}+}-\frac{1}{6}\bar{\eta}^{\bullet}\gamma^{\mu}\tilde{\hat{Q}}_{\mu}\eta_{\bullet}-\frac{i}{2}\left<g_{\hat{a}\bar{\hat{b}}}\right>\bar{\lambda}_+^{\bar{\hat{b}}}\gamma^{\mu}\tilde{\hat{\Gamma}}^{\hat{a}}_{k\mu}\lambda^{k+}\\
\nonumber &-\frac{i}{3}\left<\partial_N \mathcal{K}\right>\bar{\eta}^{\bullet}\gamma^{\mu}\tilde{\hat{\Gamma}}^{N}_{k\mu}\lambda^{k+}+(+\rightarrow -)\biggr\} \\
&\nonumber +\biggl\{ \left<\frac{1}{4b^0}g_{\hat{a}\bar{\hat{b}}}\right>\partial_{\mu}b^1\bar{\lambda}^{\bar{\hat{b}}}_+\gamma^{\mu}\lambda^{\hat{a}+}+\left<\frac{1}{18b^0}\right>\partial_{\mu}b^1\bar{\eta}^{\bullet}\gamma^{\mu}\eta_{\bullet}-(+\rightarrow -)\biggr\} \\
&\nonumber +\left<\frac{1}{4(b^0)^2}g_{\hat{a}\bar{\hat{b}}}\right> \frac{e_3M^N}{{\rm{Re}}\lambda (|m_2|+|m_1|)}B_{\mu}\left(e^{-i\theta}\bar{\lambda}^{\bar{\hat{b}}}_{+}\gamma^{\mu}\lambda^{\hat{a}-}+e^{i\theta}\bar{\lambda}^{\bar{\hat{b}}}_{-}\gamma^{\mu}\lambda^{\hat{a}+}\right)\\
&\nonumber -\left<\frac{i}{4(b^0)^2}g_{\hat{a}\bar{\hat{b}}}\right> \frac{e_3M^N}{{\rm{Re}}\lambda (|m_2|-|m_1|)}B'_{\mu}\left(e^{-i\theta}\bar{\lambda}^{\bar{\hat{b}}}_{+}\gamma^{\mu}\lambda^{\hat{a}-}-e^{i\theta}\bar{\lambda}^{\bar{\hat{b}}}_{-}\gamma^{\mu}\lambda^{\hat{a}+}\right)\\
&\nonumber +\left<\frac{1}{18(b^0)^2}\right> \frac{e_3M^N}{{\rm{Re}}\lambda (|m_2|+|m_1|)}B_{\mu}\left(e^{-i\theta}\bar{\eta}^{\bullet}\gamma^{\mu}\tilde{\eta}_{\bullet}+e^{i\theta}\bar{\tilde{\eta}}^{\bullet}\gamma^{\mu}\eta_{\bullet}\right)\\
& -\left<\frac{i}{18(b^0)^2}\right> \frac{e_3M^N}{{\rm{Re}}\lambda (|m_2|-|m_1|)}B'_{\mu}\left(e^{-i\theta}\bar{\eta}^{\bullet}\gamma^{\mu}\tilde{\eta}_{\bullet}-e^{i\theta}\bar{\tilde{\eta}}^{\bullet}\gamma^{\mu}\eta_{\bullet}\right)+{\rm{h.c.}},\label{Lint4}
\end{align}
where we have defined the fluctuations as $\hat{Q}_{\mu}=\langle\hat{Q}_{\mu}\rangle+\tilde{\hat{Q}}_{\mu}$ and $\hat{\Gamma}_{i\mu}^k=\langle\hat{\Gamma}_{i\mu}^k\rangle+\tilde{\hat{\Gamma}}_{i\mu}^k$. They are the functions of $\tilde{z}^i$.

\end{appendix}


\end{document}